\documentclass[english,floatfix,superscriptaddress,nofootinbib,prd]{revtex4}
\usepackage[paperwidth=210mm,paperheight=297mm,centering,hmargin=2cm,vmargin=2cm]{geometry}
\usepackage[utf8]{inputenc}
\usepackage[T1]{fontenc}
\usepackage{lmodern}
\usepackage{amsmath}
\usepackage{amssymb}
\usepackage{mathtools}
\usepackage{graphicx}
\usepackage{subcaption}
\usepackage{esint}
\usepackage{dcolumn}
\usepackage{babel}
\usepackage{xcolor} 
\usepackage{csquotes}
\usepackage{marvosym}
\usepackage{hyperref}
\usepackage[overload]{textcase}
\usepackage[symbol,hang,flushmargin]{footmisc}

\DefineFNsymbolsTM{otherfnsymbols}{%
  \Radioactivity	\circ
  \Yinyang   \mathsection
  \Biohazard    \dagger
  \Bat \mathsection
}%

\setfnsymbol{otherfnsymbols}

\newenvironment{seqn}{\equation\aligned}{\endaligned\endequation}

\newcommand{\be}{\begin{seqn}}
\newcommand{\ee}{\end{seqn}}

\newcommand{\bea}{\begin{eqnarray}}
\newcommand{\eea}{\end{eqnarray}}

\newenvironment{arabicfootnotes}
  {\par\edef\savedfootnotenumber{\number\value{footnote}}
   
   \setcounter{footnote}{0}}
  {\par\setcounter{footnote}{\savedfootnotenumber}}

\begin{document}
%
%
%
%
%
%

\title{Massless Charged Particles Tunneling  Radiation from a RN-dS Horizon\\ and the Linear and Quadratic GUP}


\author{Elias~C.~Vagenas}
\email{elias.vagenas@ku.edu.kw}
\affiliation{Theoretical Physics Group, Department of Physics, Kuwait University, P.O. Box 5969, Safat 13060, Kuwait.}

\author{Ahmed~Farag~Ali}
\email{ahmed.ali@fsc.bu.edu.eg}
\affiliation{Department of Physics, Faculty of Science, Benha University, Benha, 13518, Egypt.}

\author{Mohammed~Hemeda}
\email{mhemeda@sci.asu.edu.eg}
\affiliation{Department of Mathematics, Faculty of Science, Ain Shams University, 11566, Cairo, Egypt.}

\author{Hassan~Alshal}
\email{halshal@sci.cu.edu.eg}
\affiliation{Department of Physics, Faculty of Science, Cairo University, Giza, 12613, Egypt.}
\affiliation{Department of Physics, University of Miami, Coral Gables, FL 33146, USA.}

\begin{abstract}
\begin{center}
\textbf{ABSTRACT}    
\end{center}

\par\noindent
In this paper, we investigate  the massless Reissner-Nordstrom de Sitter metric in the context of minimal length scenarios. We prove  not only the confinement of the energy density of massless charged particles, both fermions and bosons, but also their ability to tunnel through the cosmological horizon. These massless particles might be interacting  with Dirac sea and in this case they will  appear outside the cosmological horizon in the context of dS/CFT holography. This result may formulate a fundamental reason for the expansion of the Dirac sea. Therefore, a spacetime Big Crunch may occur.

\end{abstract}
\maketitle
\begin{arabicfootnotes}
%
%
%
%
%
%
%
%
%
\vspace*{-0.5cm}
\section{Introduction}
%
%
%
%
\par\noindent
The existence of a minimum measurable length which has been predicted by various quantum gravity theories, such as perturbative string theory and black hole physics, opened the gate to modify the standard Heisenberg Uncertainty Principle (HUP) into the so-called  Generalized Uncertainty Principle (GUP) \cite{Veneziano:1986zf, Gross:1987ar, Amati:1988tn, Konishi:1989wk, Maggiore:1993rv, Garay:1994en, Scardigli:1999jh, Adler:2001vs}. Similar modification has been predicted by Doubly Special Relativity (DSR) theories \cite{Magueijo:2001cr, Magueijo:2004vv, Cortes:2004qn}. Based on a series of papers \cite{Das:2008kaa, Das:2009hs, Ali:2009zq, Ali:2010yn, Das:2010sj, Das:2010zf, Ali:2011fa,Ali:2011ap,Ali:2012hp}, extra linear and quadratic terms in momentum were introduced to the HUP, henceforth, named Linear and Quadratic GUP (LQGUP), in the form of 
%
%
%
%
%
\be\label{DeltaX1}
\Delta x \geq \hbar \left(\frac{1}{\Delta p} - \alpha + 4 \alpha^2 ~\Delta p\right)~,
\ee
\par\noindent
where $\alpha=\alpha_0 \ell_p /\hbar$ and $\ell_p = \sqrt{G\hbar/c^{3}}=\hbar /m_p c$. Consequently, the commutation relations which are compatible with String Theory, DSR, and consistent with the commutators of phase space coordinates $[x_i, x_j] = [p_i, p_j] = 0$ (via Jacobi Identity) take the form 
\cite{Ali:2009zq}
\be\label{GUPcomm1}
\left[x_i, p_j\right]=i\hbar\left[\delta_{ij}-\alpha\left(\delta_{ij}~p+\frac{p_i p_j}{p}\right)+\alpha^2\left(\delta_{ij}~p^2 +3p_i p_j\right)\right]
\ee
\par\noindent
which, for the case  $i=j$ and spacetime dimension D=1, can be written as 
\be\label{GUPcomm2}
\left[\hat{x}, \hat{p}\right]=i\hbar\left(1 - 2\alpha p + 4\alpha^2 p^2\right)~.
\ee
\par\noindent
Being motivated by the new findings in a series of papers by the authors \cite{Vagenas:2018pez, Vagenas:2019rai,Vagenas:2019wzd}, and, especially, by the existence of massless charged particles and a naked singularity in a massless Reissner-Nordström-de Sitter-like (RNdS) spacetime in the context of LQGUP, we are interested in studying the tunneling phenomena through the cosmological horizon of the RNdS spacetime.
The massless black holes are of physical importance. They appear in the presence of tachyonic fields \cite{Chan:1994qa}. In addition, they appear as a remedy for conifold singularities of the moduli space with the vacua of type-II string theories, known as \emph{the massless charged Ramond-Ramond black holes} \cite{Strominger:1995cz}. This type of black holes emerges when extremal black holes become massless. Furthermore, there is another type of four-dimensional massless black holes that corresponds to supersymmetric black holes with BPS solutions of effective string vacua that become massless \cite{Behrndt:1995tr}. 
This last type corresponds to massless \emph{diholes} \cite{Emparan:1999au} under the condition that supersymmetry is unbroken since Penrose cosmic censorship is not applicable to microscopic objects, and it is supersymmetry that will act as a cosmic censor for the naked singularity of the microscopic objects \cite{Ortin:1996fj,Ortin:1996nd}.\\
\par\noindent
The importance of the positive cosmological constant in the metric of the RNdS spacetime comes from the fact that such term makes the spacetime highly dynamical, especially for the diholes and the multi-black hole solutions of the Einstein-Maxwell equation \cite{Kastor:1992nn}. Such system in a de Sitter background is restricted to be an extremal black hole as long as there are massive particles in the vicinity of the system, i.e., the black hole would practice a \emph{quasistatic} discharge. Meanwhile, if there exists at least one very light, or even massless, charged particle, then the black hole becomes \emph{superextremal}, i.e., the black hole would practice an \emph{adiabatic} discharge until a Big Crunch takes place and the singularity becomes \emph{naked} \footnote{Compare the ``shark fin'' in Figure 1 of Ref. \cite{Montero:2019ekk} with the reflected one in Figure 14 of Ref. \cite{Fernando:2014wma}.} \cite{Montero:2019ekk}. It is conjectured in Ref. \cite{Montero:2019ekk} that such light charged particles are forbidden to exist due to the weak gravity conjecture that is associated with the Penrose cosmic censorship. This conjecture is proposed to resolve an apparent conundrum in the very recently proposed universal relation between extremality and entropy \cite{Goon:2019faz} in which extremal ``parental'' black holes are allowed to split, \emph{ad infinitum}, into superextremal and subextremal ``siblings'' \cite{ArkaniHamed:2006dz}. Since the weak gravity conjecture is directly related to the second law of thermodynamics \cite{Cheung:2018cwt}, and as we proved that within the LQGUP framework the existence of massless charged particles is not contradicted with the second law of thermodynamics \cite{Vagenas:2019rai}, then we are still allowed to study the massless RNdS spacetime as the ultimate superextremal charged black hole. In particular, we are interested in studying the tunneling of the massless charged particles from the massless RNdS spacetime found in Ref. \cite{Vagenas:2019rai} in order to answer an old question about how a hypothetical observer outside of the cosmological horizon would realize the tunneling according to the dS/CFT holography \cite{Strominger:2001pn} despite the fact that the question originally aimed at the tunneling in extremal black holes \cite{Medved:2002zj}.\\
\par\noindent
The rest of the paper is organized as follows. We mainly follow the analysis of Ref. \cite{Li:2016yfd} to see how to study the tunneling of massless charged particles from the cosmological horizon of the massless RNdS spacetime. In particular, we utilize the analysis of Ref. \cite{Li:2016yfd} in order to avoid dealing with a fractional Laplacian that would have appeared if we combined the standard tunneling approach \cite{Kraus:1994by,Parikh:1999mf} with LQGUP rather than the more conventional quadratic GUP \cite{Nozari:2005ix,Chen:2013tha,Banerjee:2008cf,Moayedi:2010vp}. Despite the difficulties in the mathematical computations, this proposed way is applicable to study the tunneling for any other black hole spacetime within the context of LQGUP. 
In section \textbf{II}, in the context of LQGUP we study the effect of the corresponding weight factor on the de Broglie wave length of the tunneling fields. 
Moreover, we comment on how the LQGUP introduces an effective gravitational field strength in Colella, Overhauser and Werner experiment. 
%
In section \textbf{III}, we calculate the tidal force corresponding to the massless RNdS spacetime in order to obtain the characteristic momentum related to the effective Newton's constant. 
In section \textbf{IV}, we propose a general methodology for calculating the tunneling through the cosmological horizon of the massless RNdS spacetime within the framework of LQGUP. The principle of wave-particle duality and the heuristic methodology we adopt for the tunneling from the massless RNdS spacetime stay the same for the tunneling out of any black hole spacetime within the context of LQGUP. 
We face some difficulties in calculating this particular tunneling out of the massless RNdS spacetime within the context of LQGUP despite the fact that the principle is correct. Therefore, in section \textbf{V} we consider studying the effect of the LQGUP, in the massless RNdS spacetime, on the tunneling of \emph{spin} fields.
Finally, in section \textbf{VI} we comment on the relationship between the previous findings and the question about how a hypothetical observer outside of the cosmological horizon would recognize the tunneling phenomena through the cosmological horizon.

%
%
%
%
\section{LQGUP-modified wave-particle duality}
%
%
%
%
\par\noindent
In Ref. \cite{Li:2016yfd}, the authors adopt the quadratic GUP so the momentum wavefunction, and, consequently, the de Broglie wavelength is deduced from the HUP commutation relation. The adopted method  follows the inverse proportionality between the distance and the energy $\Delta x \sim 1/\Delta E$ \cite{Garay:1994en}. As we adopt here the LQGUP, we need to consider a new weight factor from the commutation relation given by Eq. (\ref{GUPcomm2}) in order to calculate the wavelength. So, the momentum-dependent factor z(p) can be defined as
\be
\label{xP1}
z(p) =\frac{1}{i\hbar}
\left[\hat{x}, \hat{p}\right]=\left(1 - 2\alpha p + 4\alpha^2p^2\right)~.
\ee
In the case of the quadratic GUP, in order for a lower bound on the measurable length to be ensured, the average value of its $z=z(p)$ was obtained utilizing the quantity $1/(1+\beta p^2)$ as a weight factor \cite{Kempf:1994su}. For the case of the LQGUP, the weight factor  will be, up to ${\cal O}(\alpha^{2})$,  \cite{Vagenas:2019wzd}
\be\label{zWeight}
\frac{1}{\left(1-\alpha p +\frac{3}{2} \alpha^2 p^2\right)^2} \approx \frac{1}{(1-2\alpha p+4\alpha^2 p^2)} =\frac{1}{z(p)}~.
\ee
Hence, the average value of $z=z(p)$ becomes
\be\label{DeltaX2}
\Delta x \geq \frac{z \hbar}{\Delta p} \geq \ell_p~,
\ee
which confirms spacetime discreteness \cite{Ali:2009zq, Deb:2016psq,Das:2020ujn}. In order to find the relationship between the wavelength and the momentum in the context of LQGUP, the analysis of the Hilbert space representation as followed in the context of quadratic GUP \cite{Kempf:1994su} needs to be replaced with another similar analysis in order to work for the case of LQGUP.\\
\null \\
In the momentum space, the eigenvalue problem for the position operator in the context of  LQGUP  reads 
\be\label{EigenEq}
i\hbar \left(1-2 \alpha p + 4 \alpha^2 p^2 \right) \partial_p \psi_\lambda (p)=\lambda \psi_\lambda (p)
\ee
with the corresponding eigenvectors to be of the form
\be\label{EignSoln1}
\psi_\lambda (p) = C \exp\left[\frac{-i\lambda \tan^{-1}\left(\frac{4 \alpha p-1}{\sqrt{3}}\right)}{\hbar \sqrt{3} \alpha}\right]
\ee
and the normalization factor, i.e., $C$, to satisfy the condtion
\be\label{Norm}
1 = 2~||C||^2 \int_{\frac{1}{4\alpha}}^{\infty} \frac{dp}{1-2 \alpha p + 4 \alpha^2 p^2} = ||C||^2 \frac{\pi}{\sqrt{3} \alpha}~.
\ee
Hence, the final normalized eigenvectors are of the form
\be\label{EignSoln2}
\psi_\lambda (p) = \sqrt{\frac{\sqrt{3} \alpha}{\pi}} \exp\left[\frac{-i\lambda \tan^{-1}\left(\frac{4 \alpha p-1}{\sqrt{3}}\right)}{\hbar \sqrt{3} \alpha}\right]~,
\ee
and the wavelengths as functions of the momentum to read
\be\label{lambda}
\lambda = \frac{2 \pi \hbar \sqrt{3} \alpha}{\tan^{-1} \left[\frac{4 \alpha p-1}{\sqrt{3}}\right]}~.
\ee
It should be pointed out that Eq. (\ref{lambda}) is very similar to Eq. (5) in Ref. \cite{Li:2016yfd}, where our $\alpha = \alpha_{0}\ell_p$, while their $\ell_p^2\alpha$ is equal to the parameter $\beta$ in Ref. \cite{Kempf:1994su}.
We can check that Eq. (\ref{lambda}) satisfies the following relation
\be\label{DiffEq}
\frac{d}{dp} \left( \frac{2 \pi}{\lambda}\right) = \frac{\sqrt{3}}{4} \frac{\hbar^{-1}}{\alpha} z^{-1} (p)~.
\ee
The last result, namely Eq. (\ref{DiffEq}), confirms that from hereon our approach and calculations should follow those in Ref. \cite{Li:2016yfd}, and the LQGUP is not fundamentally different from the quadratic GUP \footnote{Cf. Figure 1 in Ref. \cite{Vagenas:2019wzd}.}. Yet, we still need to repeat the calculations in Ref. \cite{Li:2016yfd} within the context of LQGUP rather than quadratic GUP as the quadratic GUP was claimed to prohibit the formation of the massless RNdS \cite{Xiang:2005wg}, which is not the case within the context of  LQGUP \cite{Vagenas:2019rai}. In addition, the last result can be explained as a differential equation $\displaystyle{\frac{d\hat{k}}{d\hat{p}} = \hbar^{-1} z^{-1}~}$, where $\hat{k}=k(\hat{p})$ is a function of the momentum operator, in the same way it is explained in Ref. \cite{Li:2016yfd}. However, we replace the $z(p)$ of the quadratic GUP with our $z(p)$ for the case of  LQGUP as given in Eq. (\ref{xP1}). At the end, the momentum eigenstate is the same as $\psi_p = \exp(ikx)$.\\
\par\noindent
It is worth noting that in comparison with the arctan function in Eq. (\ref{EignSoln2}), earlier attempts to introduce linear GUP \cite{Maggiore:1993kv} are associated with an arcsinh function as in the Eq. (8) in Ref. \cite{Xiang:2007sh}. The linear GUP is endowed with a weight factor like that appears in one of the very early attempts  to introduce a quantum gravity theory in 1938 by Max Born \cite{Born:1938} (see Eq. (7) therein). The point is that for different uncertainty principle physics, whether it is HUP, or any kind of the GUP, the wavelength $\lambda$ also varies depending on the weight factor we apply. However, the eigenfunctions of LQGUP are closer to the ones of the quadratic GUP than to the ones of the linear GUP, which is expected from the convergent behavior in both quadratic GUP and LQGUP in Ref. \cite{Vagenas:2019wzd}. 
%
%
%
%
%
%
%
%
%
%
\par\noindent
The quantum pattern of two neutron beams induced by gravity is discussed by Colella, Overhauser, and Werner (COW) \cite{Colella:1975dq}. A nice summary of the phase shift between the two beams associated with the COW analysis is presented in Ref. \cite{Li:2016yfd}. The phase shift is a linear function of the gravitational acceleration. Heuristically, the same phase shift is also a function of the change of the wave number with respect to the momentum. Therefore, we can relate the gravitational acceleration to the previous result in Eq. (\ref{DiffEq}), where in our case the $z(p)$ is that of LQGUP as given in Eq. (\ref{xP1}). Therefore, the phase shift becomes
\be\label{NewPhasShift2}
\Delta \phi^\prime &\approx l\frac{dk}{dp} \Delta{p} \\ &= \hbar^{-1} z^{-1} l \Delta p
\ee
where $l$ is the horizontal path of the beam that is perpendicular to the gravitational acceleration. At the same time, the conservation of energy and the equation between potential and kinetic energy imply that the phase shift becomes
\be\label{NewPhasShift3}
\Delta \phi^\prime = \frac{mgyl}{z\hbar v} = \frac{mg^\prime A}{\hbar v}~,
\ee
where $v$ is the beam velocity, $g^\prime = g/z$, $A = yl$, and $y$ is the beam path parallel to the gravitational acceleration. In the context of HUP, $z = 1$ and, hence, Eq. (\ref{NewPhasShift3}) returns to $\displaystyle{\Delta \phi = \frac{mgA}{\hbar v}}$, which matches with the earlier result expected from the conventional quantum theory without considering strong gravitational effects.\\
\par\noindent
As in the previous findings of $\Delta \phi$ and Eq. (\ref{NewPhasShift3}), the gravitationally corrected phase shift is almost the same as that of the HUP, except for the  momentum-dependent factor $z(p)$, which is a function of the momentum in the presence of the strong gravity. This $z(p)$ suggests replacing $g$ with a corrected $g^\prime=g/z(p)$ 
due to the GUP effect, whether its linear, quadratic, or LQGUP, on the COW experiment. The experiment itself becomes equivalent to how the propagation of those two neutron beams in a strong gravitational field  is characterized by an effective gravitational field strength $g^\prime$. 
Consequently, this demands a modification for the Newton's constant, i.e., from $G$ to $G^\prime$, in the context of LQGUP, after we choose its corresponding $z(p)$. Therefore, we can introduce another effective physical quantity $G^\prime$ as the \emph{effective Newton's constant}. This corollary comes as the effective field strength is written as
\be\label{}
g^\prime = g/z = \frac{GM}{zR^2} = \frac{G^\prime M}{R^2}
\ee
where $R$ is the radius of the gravity source, and $G^\prime = G/z$ as in Ref. \cite{Xiang:2013sza}. The next step is to restrict our analysis to $z=z(p)\equiv z_{\text{LQGUP}}$.\\
\par\noindent
It is noteworthy that the COW experiment has been discussed within some newly introduced DSR-GUP regime \cite{Farahani:2020ctt}. Despite the fact that it is different from the LQGUP regime we adopt here, it agrees with our analysis that the effective Newton's constant is characterized by the weight factor corresponding to the chosen GUP. Furthermore, it agrees  with our findings that the wavelength is a tan or an arctan function of the momentum. This is because the DSR-GUP, introduced in Ref. \cite{Chung:2018btu}, is in fact a function of $p$ and $p^2$ similar to the LQGUP. Here, we work in the context of  LQGUP because in this context the massless RNdS spacetime is consistent with the second law of thermodynamics. Now, we are interested in studying the tunneling through the cosmological horizon of the massless RNdS spacetime as presented in the next sections.
%
%
%
%
\section{Gravitational tidal force and the characteristic momentum}
%
%
%
%
\par\noindent
Now we consider the massless charged particles throughout the massless RNdS-like spacetime whose metric is given by
\be\label{}
ds^2 = -\left( 1 + \frac{Q^2}{r^2} - \frac{\Lambda r^2}{3}\right) c^2 dt^2 + \left( 1 + \frac{Q^2}{r^2} - \frac{\Lambda r^2}{3}\right)^{-1} dr^2 + r^2 d \Omega
\ee
where $\Lambda = 8\pi \frac{G}{c^2} \rho_{vac}$ is the cosmological constant in the standard HUP, and $\rho_{vac}$ is the vacuum density. However, in LQGUP, this $\Lambda$ becomes $\Lambda^\prime = 8\pi \frac{G^\prime}{c^2} \rho_{vac}$ with $G^\prime = G/z_{\text{LQGUP}}$. Thus, the massless charged RNdS-like spacetime metric within LQGUP will take the following form
\be\label{RNdS}
ds^2 = -\left( 1 + \frac{Q^2}{r^2} - \frac{\Lambda^\prime r^2}{3}\right) c^2 dt^2 + \left( 1 + \frac{Q^2}{r^2} - \frac{\Lambda^\prime r^2}{3}\right)^{-1} dr^2 + r^2 d \Omega~,
\ee
with the corresponding LQGUP-modified Hawking temperature to be of the form \cite{Vagenas:2019rai}
\be\label{}
T_H^\prime &= \frac{\hbar\left(\beta  - \sqrt{\beta^2  - 16 \alpha^2}\right)}{8\alpha^2} \\ &=
 \frac{2\,\hbar}{\left(\beta  + \sqrt{\beta^2 - 16 \alpha^2}\right)}
\ee
\par\noindent
with $\beta$ to be the inverse unmodified Hawking temperature, i.e., $\beta = \frac{\hbar}{T_H}$.\\
As in Ref. \cite{Li:2016yfd}, the momentum $\Delta p$ associated with tidal force in a curved spacetime is characterized by the Riemann tensor $R_{\rho\lambda\mu\nu}$ \cite{Crispino:2016pnv}. One of the non-vanishing independent components of the Riemann tensor, along the tangent $u^\mu$, in any spherically symmetric spacetime, including the massless RNdS spacetime, is given by \cite{Crispino:2016pnv, Narlikar:1986jv}
\be\label{}
f^r = R^r_{trt} u^t u^r u^t &= \frac{1}{9r^6} \left(3r^2 - \Lambda r^4 + 3Q^2\right)\left(9Q^2 - \Lambda r^4\right) \\&= \left(1 + \frac{Q^2}{r^2} - \frac{\Lambda}{3}r^2\right)\left(\frac{3Q^2}{r^4} - \frac{\Lambda}{3}\right) \\ &= g^{rr} h(\Lambda, Q, r)~.
\ee
Thus, its covariant form reads
\be\label{Fr}
f_r = h(\Lambda, Q, r) = \frac{3Q^2}{r^4} - \frac{\Lambda}{3}.
\ee
\par\noindent
For a pair of virtual particles with energy $\Delta E$ and  separated by a distance $\Delta x$ with $\Delta t$ as the life-time of the virtual particles, the uncertainty in the momentum due to the tidal force is 
\be\label{DeltaP1}
\Delta p = F \Delta t = h(\Lambda, Q, r) \frac{\Delta E}{c^2} \Delta t \Delta x~.
\ee
Therefore, the geodesic deviation equation, derived from Riemann tensor, defines the tidal force as
\be\label{}
F = h(\Lambda, Q, r) \frac{\Delta E}{c^2} \Delta x~.
\ee
When the virtual particles experience enough tidal force and become real, the physical observability requires $\Delta p \Delta x \geq \hbar$, $\Delta E \Delta t \geq \hbar$. Thus, we infer from Eq. (\ref{DeltaP1}) and Eq. (\ref{Fr}) that 
\be\label{DeltaP2}
(\Delta p)^2 &\geq \frac{\hbar \Delta p}{\Delta x} = \frac{\hbar F}{\Delta x} \Delta t \\ &= \frac{\hbar}{c^2}\Delta E \Delta t \left(\frac{3Q^2}{r^4} - \frac{\Lambda}{3}\right).
\ee
The last inequality suggests the existence of a characteristic momentum as
\be\label{}
\Delta p_m \approx \sqrt{\frac{3Q^2}{r^4} - \frac{\Lambda}{3}}~.
\ee
Now we see how the minimal momentum of the realized particles, produced from the quantum fields in the massless RNdS spacetime, defines the characteristic scale of the system. 
Thus, Eq. (\ref{DeltaP2}) yields the momentum uncertainty of the real particles, which can be observed by a photon with energy $\Delta E$,
\be\label{}
\Delta \widetilde{p} \geq \sqrt{\left(\frac{3Q^2}{r^4} - \frac{\Lambda}{3} \right) \frac{\hbar^2}{c^2}}, 
\ee
where $3Q^2/r^4 > \Lambda/3$ and the energy-time uncertainty is considered.\\
\par\noindent
Next we employ the identified characteristic scale in Eq. (\ref{DeltaP2}) to obtain an effective Newton's constant as
\be\label{}
G^\prime = \frac{G}{1 - \frac{2\alpha}{\hbar}\Delta \widetilde{p} + \frac{4\alpha^2}{\hbar^2}\Delta \widetilde{p}^2}~.
\ee
At this point a couple of comments are in order. First, we emphasize that the last result is true for any other static spherically symmetric spacetime. Therefore, it is enough to find the specific correction factor $h=h(M, Q, J, \Lambda, r)$, determined by the black hole characteristic ``hairs'', for the spacetime under study.
Second, one may argue that for $r$ large enough there will be some values for the set of $Q$ and $\Lambda$ for which $3Q^2/r^4 <\Lambda/3$ and thus the last three equations will be imaginary/complex. Obviously, this is true in the IR limit. However, in the same limit we have $p^2<<m^2$. Since here we consider \textit{massless} particles, then $p^2<<0$ means the momentum is allowed to be complex. This is in fact a good corollary as it serves to reintroduce the ill-defined cut-off integrals, that appear in Feynman diagrams, in complex planes. For more details on this issue, one can  see in Ref. \cite{Buchbinder:2021wzv}. For the LQGUP corrected version of those integrals, one can follow the approach in Ref. \cite{Vagenas:2019wzd}.
%
%
%
%
\section{Quantum tunneling of massless charged particles}
%
%
%
%
\par\noindent
We recall the metric of the massless charged particles throughout RNdS-like spacetime given by Eq. (\ref{RNdS}) and express it in the form \footnote{In this section, we have set $\hbar = c =1$.} \cite{Vagenas:2019rai}
\be\label{metric}
ds^2 = -\left( 1 + \Phi(r)\right) dt^2 + \left( 1 + \Phi(r)\right)^{-1} dr^2 + r^2 d\Omega^2
\ee
where the charge $Q$ in the potential $\Phi(r)$ is a function of the radius of the cosmological horizon $r_c$, which is also a function of the LQGUP corrected cosmological constant $\Lambda'$ as given below. This charge $Q$ should be the one that allows the existence of the massless charged particles as shown in Ref. \cite{Vagenas:2019rai}. So, the potential reads
\be\label{effectivemetricLambda}
\Phi(r) = \frac{Q^2(\Lambda^\prime, r_c)}{r^2} - 
\frac{\Lambda^\prime}{3}r^2
\ee 
and the corrected cosmological constant is
\be\label{lambda1}
\Lambda^\prime = \frac{\rho_{vac} ~G^\prime}{3} = \frac{\rho_{vac} ~G}{3\left[1 - 2\alpha \left(\frac{3Q^2}{r^4} - \frac{\Lambda}{3} \right)^{1/2}  + 4\alpha^2 \left(\frac{3Q^2}{r^4} - \frac{\Lambda}{3} \right) \right]}~.
\ee
Here we use $\Lambda \sim \rho_{vac}$, while $Q^2(\Lambda^\prime, r_c)$ and $r_c(\Lambda)$ can be obtained from Ref. \cite{Vagenas:2019rai} as
\be\label{Qcorrected}
Q^2(\Lambda^\prime, r_c) &= \frac{\Lambda^\prime r_c^4}{3} - r_c^2
\ee
and 
\be\label{rcosmo}
r_c &= \left(\frac{\pi}{4\alpha \Lambda} \right)+ \sqrt{\left(\frac{\pi}{4\alpha \Lambda}\right)^2 +\frac{3}{\Lambda}}~.
\ee
It should be pointed out that the cosmological constant $\Lambda$ in the above equation does not need to be LQGUP corrected because the $r_c$ is already LQGUP corrected, see Ref. \cite{Vagenas:2019rai}. Therefore, if we directly substitute Eq. (\ref{lambda1}) and Eq. (\ref{rcosmo}) in Eq. (\ref{Qcorrected}), then $\Phi$ becomes fully LQGUP corrected.\\
\par\noindent
In order to get the horizon, we need to set $g^{rr}=0$ and, thus, we obtain
\be\label{phi}
1 + \Phi(r) = 0~.
\ee
At this point, it should be stressed that 
in order for our system to have a naked singularity, 
the charge $Q(\Lambda^\prime,r_c)$ has to use for the cosmological constant the value $\Lambda^\prime$ 
as given in Eq. (\ref{lambda1}), and the value for the $r_c$ as given in Eq. (\ref{rcosmo}).
It is evident that $\Phi(r)$ has a very complicated mathematical form and, thus, an analytic solution of Eq. (\ref{phi}) is not easily obtained.\\
\par\noindent
The tunneling probability of the black hole radiation through the horizon is determined by the imaginary part of the action \cite{Parikh:1999mf, Parikh:2004ih}. This probability is restricted by the change in the Bekenstein-Hawking entropy at the WKB approximation. The tunneling probability is given by
\be\label{gamma}
\Gamma \sim \exp\left(\operatorname{Im}[S]\right)=\exp\left(\operatorname{Im}[\oint p_{r}dr]\right)=\exp\left(\operatorname{Im}[\int \limits^{r_f}_{r_i} p_{r}dr]\right)
\ee
where $r_i$ is the initial radius of the black hole at the beginning of tunneling process, and $r_f$ the final radius of tunneling. The imaginary quantity in the action is responsible for a particle tunneling through the cosmological horizon from $r_i$ to $r_f$. In addition, the action should be calculated according to Refs. \cite{Akhmedov:2006pg, Akhmedov:2006un, Akhmedova:2008dz, Akhmedov:2008ru} in order to avoid doubling the Hawking temperature and to make it invariant under canonical transformation. Therefore, to evaluate the emission rate of the spacetime, we use the Painleve type coordinate as
\cite{Parikh:1999mf,Vagenas:2002hs}
\be\label{}
\tau = t + \int \frac{\sqrt{-\Phi}}{1+\Phi} dr~.
\ee
Then, the metric given in Eq. (\ref{metric}) can be rewritten as
\be\label{effectivemetricPainleve}
ds^2 = -(1 + \Phi(r)) d\tau^2 + 2 \sqrt{-\Phi(r)} dr d\tau + dr^2 + r^2 d\Omega~.
\ee
Interestingly, Eq. (25) of Ref. \cite{Patino:2001sh}, for massive RN, is the same as  Eq. (\ref{effectivemetricPainleve}).
Moreover, there is no coordinate singularity as there is no event horizon. 
However, the massless RNdS has a geometric naked singularity. Therefore,  the massless RNdS spacetime is appropriate for the tunneling of the massless charged particles through the cosmological horizon.
When $ds^2 = 0 = d\Omega$, one can obtain the radial null geodesics as
\be\label{}
\dot{r} = \frac{dr}{d\tau} = 1 - \sqrt{-\Phi(r)}~.
\ee
Now one may try to obtain the radial momentum as
\be\label{}
p_r = \int \frac{dM}{1 - \sqrt{-\Phi(r)}} = \frac{1}{8\pi} \int \frac{-k_c dA_c - V d\Lambda}{1 - \sqrt{-\Phi(r)}}~,
\ee
where $dM \sim -k_c dA_c - V d\Lambda$ as in Ref. \cite{Vagenas:2019rai}. Unfortunately it is also hard to solve, and  $S = \oint p_r dr$ is even harder to be computed. Therefore, we have to follow a different approach.\\
%
%
%
%
\section{Tunneling of massless charged particles as spin fields}
%
%
%
%
\par\noindent
We need to check which spins are allowed to tunnel through such system, i.e., the massless RNdS spacetime. In Ref. \cite{Li:2010ad, Li:2011za, Li:2011gu, Guqiang:2016sh}, the equations of motion for the massive and massless fermions as well as bosons are considered. Here, we consider only the massless fields, with $B = (1 + \frac{Q^2}{r^2} - \frac{\Lambda}{3}r^2)$. In the massless RNdS spacetime, the mass terms tend to zero, so Eq. (10) in Ref. \cite{Li:2011za} becomes 
\be\label{rho}
\rho = - \frac{1}{r}, ~~~\gamma = \frac{-3Q^2 - \Lambda r^4}{6r^3}, ~~~\mu = - \frac{3r^2 + 3Q^2 - \Lambda r^4}{6r^3}, ~~~
\alpha = - \beta = - \frac{\cot \theta}{2\sqrt{2}r}, ~~~\Psi_2 = \frac{Q^2}{r^4}
\ee
where $\rho$, $\gamma$, $\mu$, $\alpha$ and $\beta$ are the spin coefficients that are obtained from the Ricci rotation coefficient, and $\Psi_2$ is the second Weyl-Newman-Penrose scalar obtained from the Weyl tensor. Eq. (\ref{rho}) shows that the RNdS metric is of Petrov type D. In principle, there is nothing that prohibits considering the tunneling of massless Fronsdal-Fang fields with arbitrary spins \cite{Fronsdal:1978rb,Fang:1978wz}. But due to the Vasiliev conjecture \cite{Prokushkin:1998bq,Vasiliev:1999ba}, we restrict the calculations of the tunneling to the fields with lower spins. In the massless RNdS spacetime, the field equations of the Weyl neutrino (s = $\frac{1}{2}$), electromagnetic (s = 1), massless Rarita-Schwinger (s = $\frac{3}{2}$) and gravitational (s = 2) fields can be combined into
\be\label{D1}
&\{[D - (2s -1)\epsilon + \bar{\epsilon} - 2s\rho - \bar{\rho}](\Delta - 2s\gamma + \mu) \\
&- [\delta + \bar{\pi} - \bar{\alpha} - (2s - 1)\beta - 2s\tau](\bar{\delta} + \pi - 2s\alpha) \\
&- (2s - 1)(s -1)\Psi_2\} \Phi_{+s} = 0~,
\ee
\be\label{D2}
&\{[\Delta + (2s -1)\gamma - \bar{\gamma} + 2s\mu + \bar{\mu}](D + 2s\epsilon - \rho) \\
&- [\bar{\delta} - \bar{\tau} + \bar{\beta} + (2s - 1)\alpha + 2s\pi](\delta - \tau + 2s\beta) \\
&- (2s - 1)(s -1)\Psi_2\} \Phi_{-s} = 0
\ee
where $D = l^\mu \partial_\mu$, $\Delta = n^\mu \partial_\mu$, $\delta = m^\mu \partial_\mu$ are the directional covariant derivative operators for each null tetrad direction \cite{Li:2007ae}. Here  the first equation, i.e.,  Eq. (\ref{D1}), is for spin states $\varpi = s$ and the second equation, i.e., (\ref{D2}), is for $\varpi = -s$. As we see, the energy of the fields depends on both the characteristics of the background and the spin \cite{Li:2000hd} due to the LQGUP corrections to the entropy we made in Ref. \cite{Vagenas:2019rai}. The mode functions of these spin fields around the RNdS black hole, $\Phi_\varpi = \exp[-i Et + i S_\varpi(r, \theta, \phi)]$, can be written at the WKB level using Eq. (\ref{rho}), Eq. (\ref{D1}), and Eq. (\ref{D2}) as
\be\label{E2}
\frac{E^2}{B} - B P^2_r - \frac{1}{r^2} P^2_\theta - \frac{1}{r^2 \sin^2\theta} (P_\phi + \varpi \cos\theta)^2 + \eta (r, \varpi) = 0
\ee
where $P_\mu = \partial_\mu S_\varpi$ are the conjugate momenta and 
\be\label{eta}
\eta(r, \varpi) = -\frac{2}{3} \Lambda \left[\varpi (2 \varpi + 3) + 1\right] + \frac{\varpi}{r^2} + \frac{s - \varpi}{r^4} \left[\frac{3 - 2s}{3} \Lambda r^4 + r^2 + (2s - 1) Q^2\right]
\ee
where $\varpi = \pm s$. 
From Eq. (\ref{E2}), we can obtain the 3-momentum of a massless particle moving in a massless RNdS background as
\be\label{EE}
E^2 = -B (P^2 + \eta)~.
\ee
This is before the GUP correction, where $P^2 = \left[ \frac{3Q^2}{r^4} - \frac{\Lambda}{3}\right]$ is the momentum obtained from tidal force.\\
\par\noindent
In order to obtain $E_{\text{LQGUP}}$, we use 
\be\label{E_LQGUP}
E_{\text{LQGUP}} \geq E (1 - 2\alpha P + 4\alpha^2 P^2)
\ee
where $E$ is given in Eq. (\ref{EE}) as in Ref. \cite{Haldar:2018zyv}. In that paper, the authors recognize LQGUP, but they use only the quadratic GUP. Therefore, once we find $E_{\text{LQGUP}}$ for every massless kind of fermions and bosons, we can use Eq. (26) of Ref. \cite{Haldar:2018zyv} to obtain
\be\label{GAMMA_LQGUP}
\Gamma_{\text{LQGUP}} \sim \exp(\text{Im}[S]) = \exp \left( \frac{4\pi E_{\text{LQGUP}}}{B^\prime (r_c)} \right)~.
\ee
\par\noindent 
The solutions for Eq. (\ref{E_LQGUP}) in the different spin cases $s = 0, \pm 1, \pm 2, \pm \frac{1}{2}, \pm \frac{3}{2}$ and for $\varpi = \pm s$ can be obtained after we combine Eq. (\ref{E2}), Eq. (\ref{eta}), and Eq. (\ref{EE}). Below we provide the graphs for the different combinations of spin $s$ and $\varpi$ of $E_{\text{LQGUP}}$. In all spin cases, the corresponding energy solutions are hard to be expressed analytically. However, we can graphically represent them in Figs. \ref{fig.1}, \ref{fig.2}, \ref{fig.3}, \ref{fig.4}, \ref{fig.5}. All the graphs are plotted at fixed $\alpha =1$. For the scalar fields ($s=0$), we notice in the two-dimensional Fig. \ref{fig.1}a that the highest $E_{\text{LQGUP}}$ is obtained when $\Lambda$ is maximum and $Q$ is minimum. For the fermionic massless fields, we notice that Figs. \ref{fig.2}c, \ref{fig.2}d and Figs. \ref{fig.4}c,  \ref{fig.4}d show $E^2_{\text{LQGUP}}$ has negative values. This means that for the cases $\varpi=-(s=\frac{1}{2})$ and $\varpi=-(s=\frac{3}{2})$ the fermionic massless fields do not tunnel out of the horizon. Similarly, we have restriction on the tunneling of the case $\varpi=-(s=1)$ of bosonic massless fields as we notice from Fig. \ref{fig.3}c and \ref{fig.3}d. However, the bosonic massless $\varpi=-(s=2)$ fields are not plagued with those negative $E^2_{\text{LQGUP}}$. By comparing the maximum value of each diagram (e,f,g,h) in each Figure, we see that the higher the spin, the higher the tunneling energy.

%
%
%
%
\section{Remarks and Conclusion}
%
%
%
%
\par\noindent
In this work, we mainly followed the methodology of Ref. \cite{Li:2016yfd} and  we succeeded to avoid dealing with a fractional Laplacian that would have appeared if we combined the standard tunneling approach 
in the context of LQGUP.
%
%
Despite the difficulties in the mathematical computations that we encountered at the end, our proposed method for the calculations is applicable to study the tunneling for any other black hole in the context of LQGUP. 
In section \textbf{II}, we studied the LQGUP and the effect of the corresponding weight factor on the de Broglie wavelength 
of the tunneling fields. 
We commented on how the LQGUP introduces an effective gravitational field strength in the COW experiment. 
%
In section \textbf{III}, we calculated the tidal force corresponding to the massless RNdS spacetime to get the characteristic momentum related to the effective Newton's constant. 
In section \textbf{IV}, we proposed to employ the general methodology of calculating the tunneling through the cosmological horizon of the massless RNdS spacetime in the framework of LQGUP. 
The principle of wave-particle duality and the heuristic methodology we adopted for the tunneling in the massless RNdS spacetime stay the same for the tunneling out of any black hole in the context of LQGUP. %
In section \textbf{V}, due to the mathematical difficulties in calculating this specific tunneling in the context of the LQGUP, we considered studying the effect of the LQGUP on the massless RNdS spacetime in the presence of tunneling \emph{spin} fields. In brief, we investigated the effect of minimal length on RNdS-like spacetime. We found that GUP stimulates a tunneling of massless charged particles which could give a fundamental reason for the expansion of Dirac sea associated with  a Big Crunch. The effect of GUP could be understood as a dark energy effect that accounts for the negative pressure in the universe and, hence, its continuous expansion. We hope to extend our results to cosmological models in the future.
\par
We did not consider the massless spin fields $s>2$ following the Vasiliev conjecture \cite{Prokushkin:1998bq,Vasiliev:1999ba} in dS/CFT \cite{Anninos:2011ui} which builds a ``tower'' of massive fields for $s>2$ on top of the massless ones with $s\leq 2$. The massive Singh-Hagen fields \cite{Singh:1974qz,Singh:1974rc}, their dual Curtright fields \cite{Curtright:1980yk}, and the massless Fronsdal-Fang fields \cite{Fronsdal:1978rb,Fang:1978wz} are chargeless in general. Our study here might be considered as a hint for a future study on higher-spin massless charged fields in some black hole spacetimes as well as in cosmological models. The reason is that assuming the validity of the Vasiliev conjecture, the higher-spin massless particles may later suffer symmetry breaking and, hence, they become massive.
One may wonder how it is possible to have a ``charged'' gravitational ($s = 2$) field. In fact, de Rham {\it et al} \cite{deRham:2014tga} proposed a ghost-free interaction theory of \emph{massive charged spin-2 fields}. This can be a clue for future study to link our massless spin-2 fields with the massive ones. Perhaps our massless charged spin-2 field suffers a symmetry breaking so that it becomes later a de Rham {\it et al} massive charged  spin-2 field.
\par
But what we care more is that our analysis of the massless RNdS spacetime emphasizes and corroborates Medved’s analysis \cite{Medved:2002zj} of how a hypothetical observer outside of the cosmological horizon would realize the tunneling on the underlying dS/CFT holography of the entire space. The special thing about the massless RNdS spacetime tunneling is the corresponding Big Crunch that is expected to happen with such system as we presented in the introduction of this work. Moreover, the quanta radiated through the cosmological horizon are massless, charged, and spinful, i.e., they should correspond to some conformal symmetry related to Fronsdal and Fang massless, i.e., gauge-invariant, fields in the de Sitter background \cite{Angelopoulos:1980wg}. Therefore, we cannot avoid Medved’s demand that the observer needs to be “specially” capable of globally observing the entire spacetime under study which is experiencing the Big Crunch.
\par
As in Medved’s analysis \cite{Medved:2002zj}, the special observer (entity?) outside of the horizon detects the massless fields with negative energy tunneling outwards. The positive-energy counterpart of the massless fields stays behind the cosmological horizon and \emph{effectively} increases the energy content of the background spacetime during the Big Crunch. In our case, the effective geometry is described in section \textbf{IV}, see Eq. (\ref{effectivemetricLambda}). The effective Newton’s constant corrects the cosmological constant in such a way that it plays the same role with the  effective mass $\omega$ in Medved analysis. Therefore, Eq. (\ref{effectivemetricPainleve}) is the metric for the ``topological'' massless RNdS spacetime. This agrees also with the fact that such spacetime system must have a naked singularity, which is what we proved before. Finally, we second Medved's statement : 
{\it ``\dots It is quite possible that there are deep connections between semi-classical thermodynamics and de Sitter holography that await to be uncovered. We hope to report progress along these lines at a future date.  \dots ''}.
\\
%
%
%
%
%
%
%
%
%
%
%
\begin{figure}[h!]
\begin{minipage}[t]{0.45\linewidth}
\captionsetup{justification=centering}
\includegraphics[width=\linewidth]{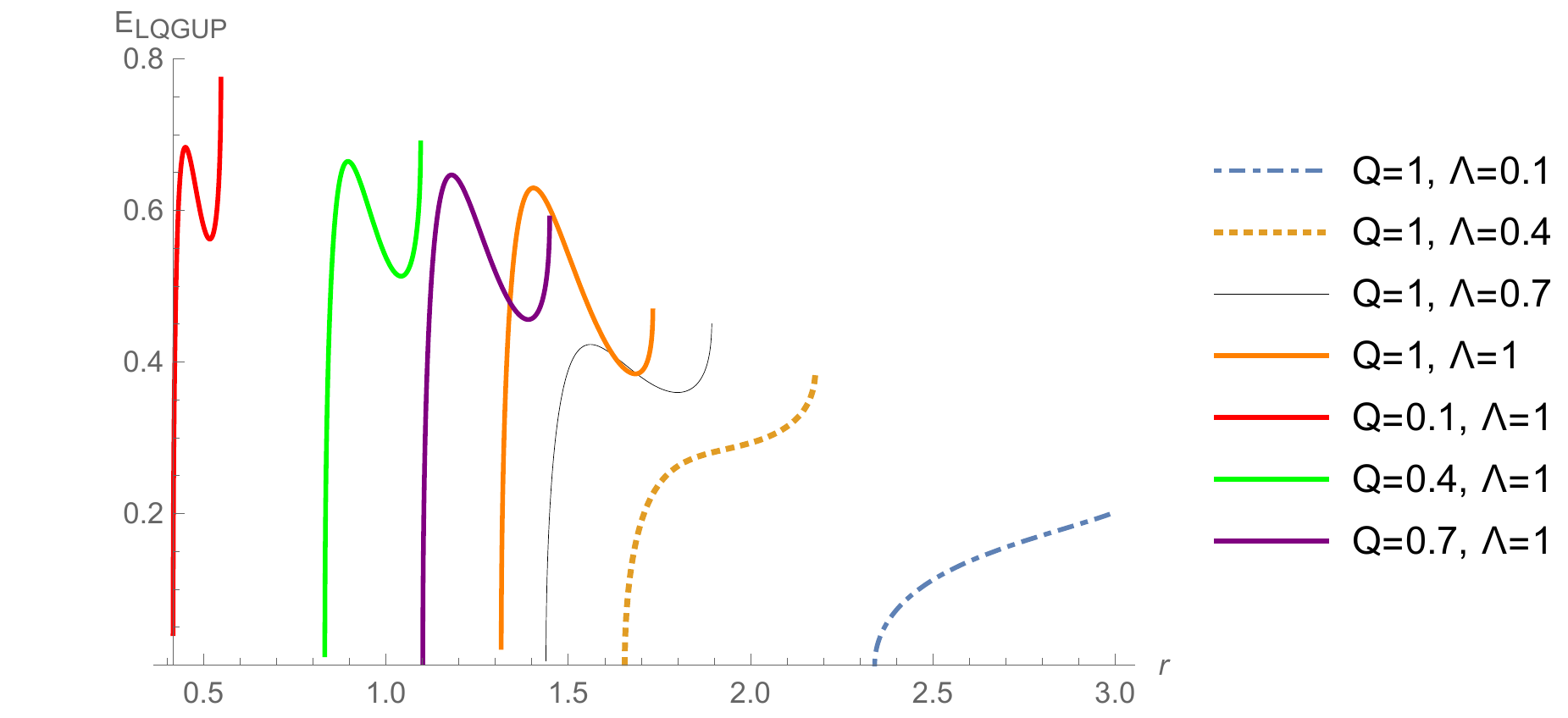}
\subcaption{ $E_{\text{LQGUP}}$ against $r$ for different $Q, \Lambda$ at $\alpha=1$.}
\end{minipage}\hfill
\begin{minipage}[t]{0.45\linewidth}
\captionsetup{justification=centering}
\includegraphics[width=\linewidth]{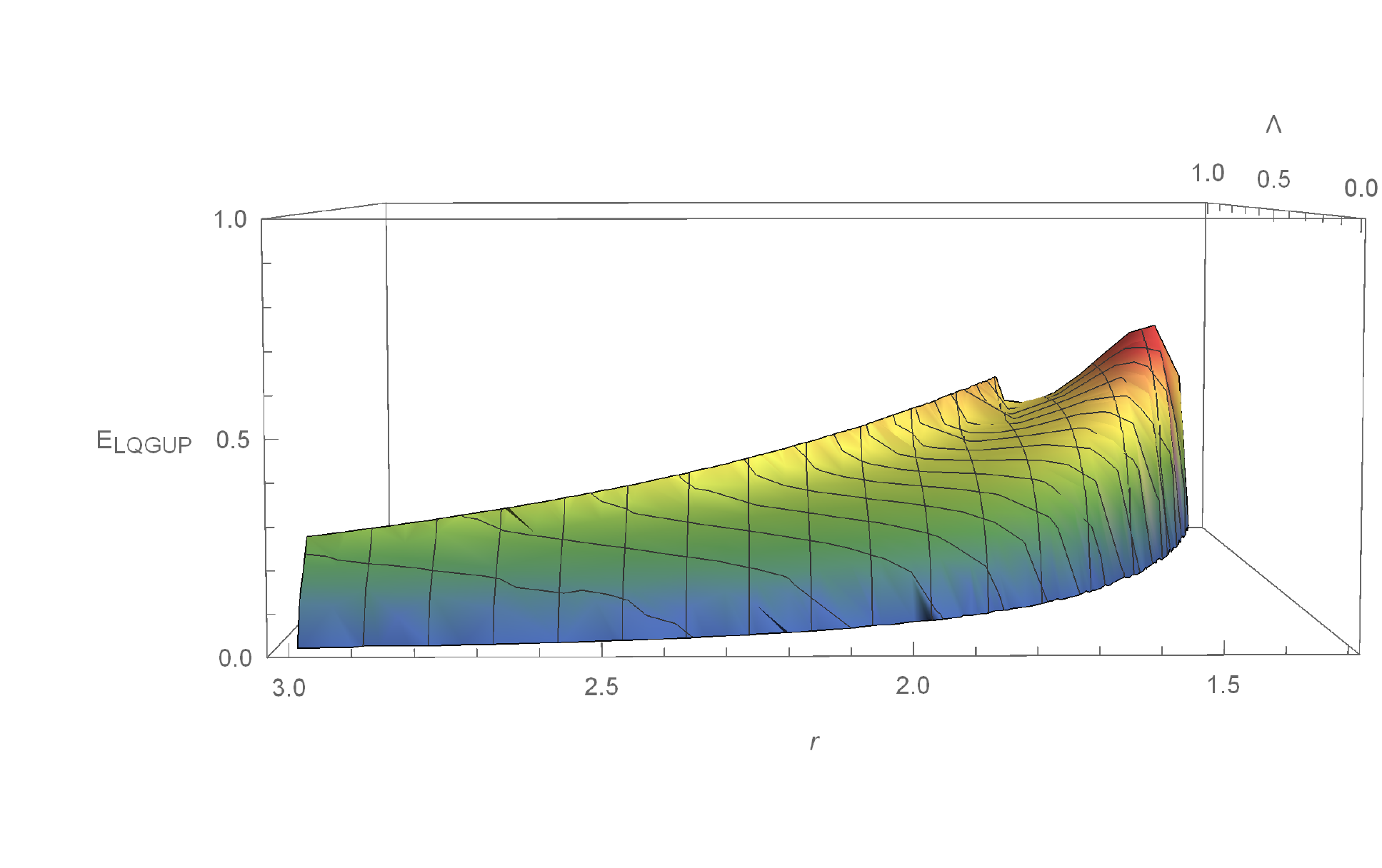}
\subcaption{ $Q=\alpha=1, \Lambda$ varies.}
\end{minipage}

\begin{minipage}[t]{0.45\linewidth}
\captionsetup{justification=centering}
\includegraphics[width=\linewidth]{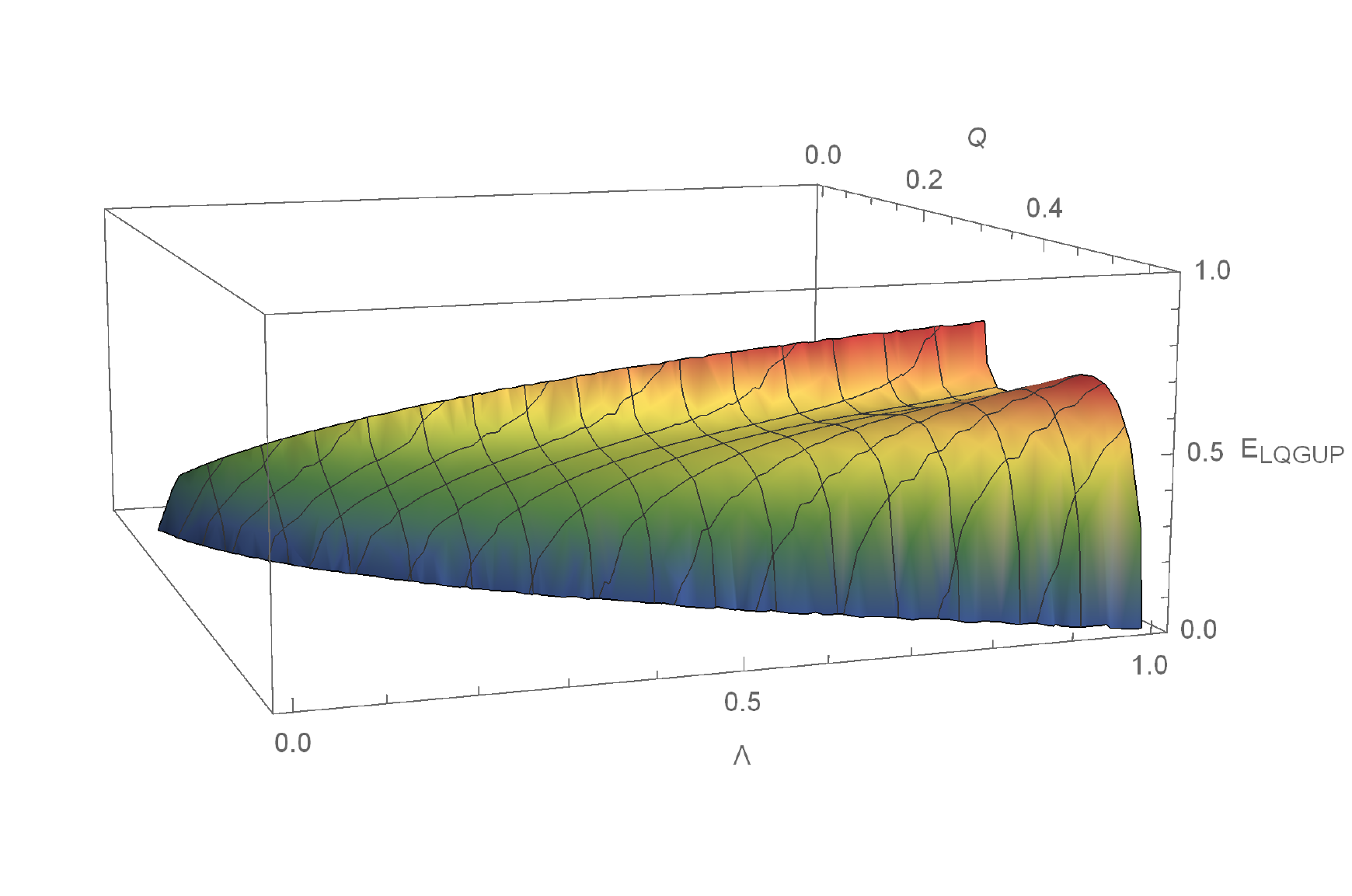}
\subcaption{ $r=\alpha=1$, $\Lambda$ and $Q$ vary.}
\end{minipage}\hfill
\begin{minipage}[t]{0.45\linewidth}
\captionsetup{justification=centering}
\includegraphics[width=\linewidth]{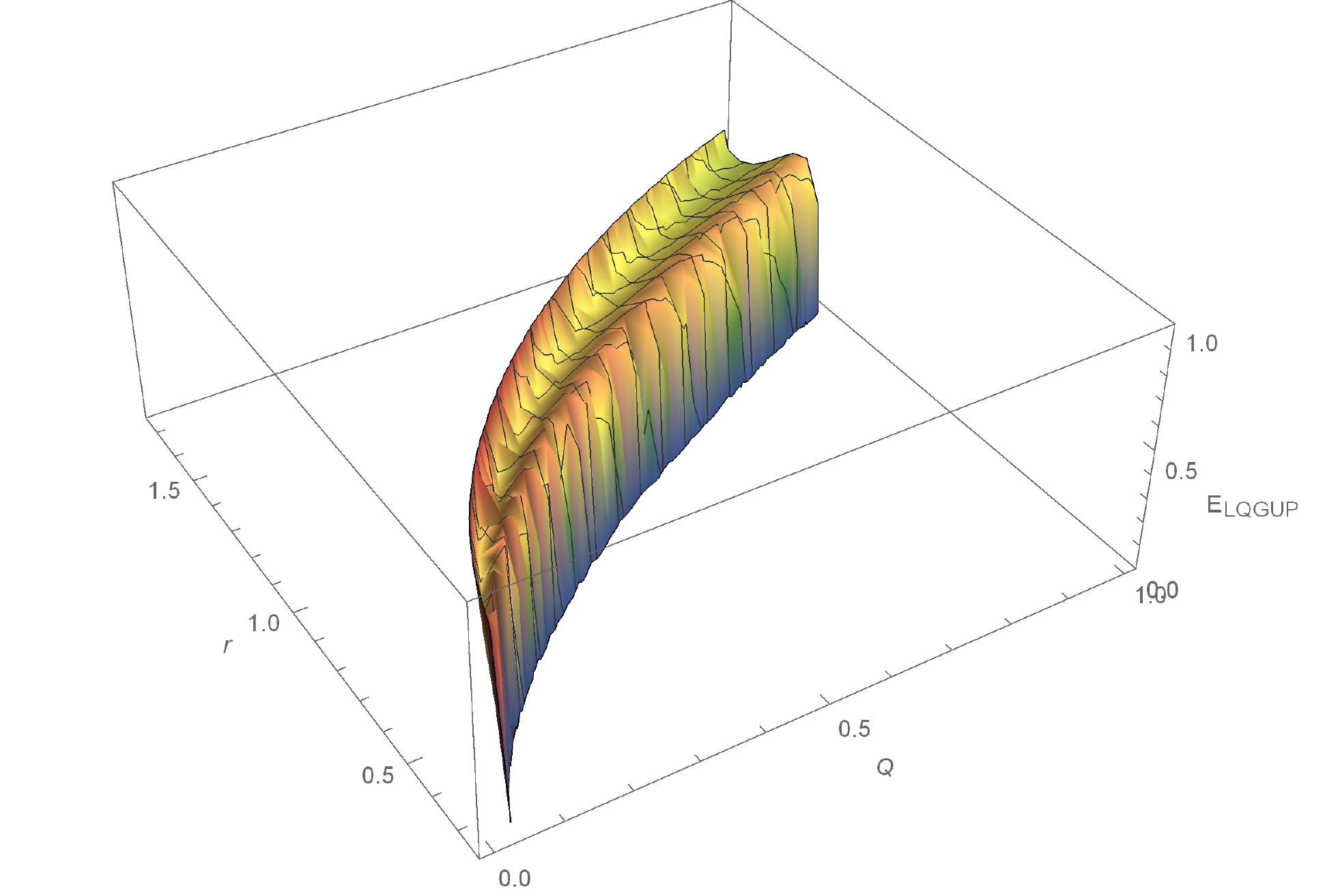}
\subcaption{ $\Lambda=\alpha=1, Q$ varies.}
\end{minipage}

\captionsetup{justification=centering}
\caption{Tunneling energy $E_{\text{LQGUP}}$ for $\varpi=s=0$ scalar fields.}\label{fig.1}
\end{figure}

\newpage


\begin{figure}[h!]
\begin{minipage}[t]{0.45\linewidth}
\captionsetup{justification=centering}
\includegraphics[width=\linewidth]{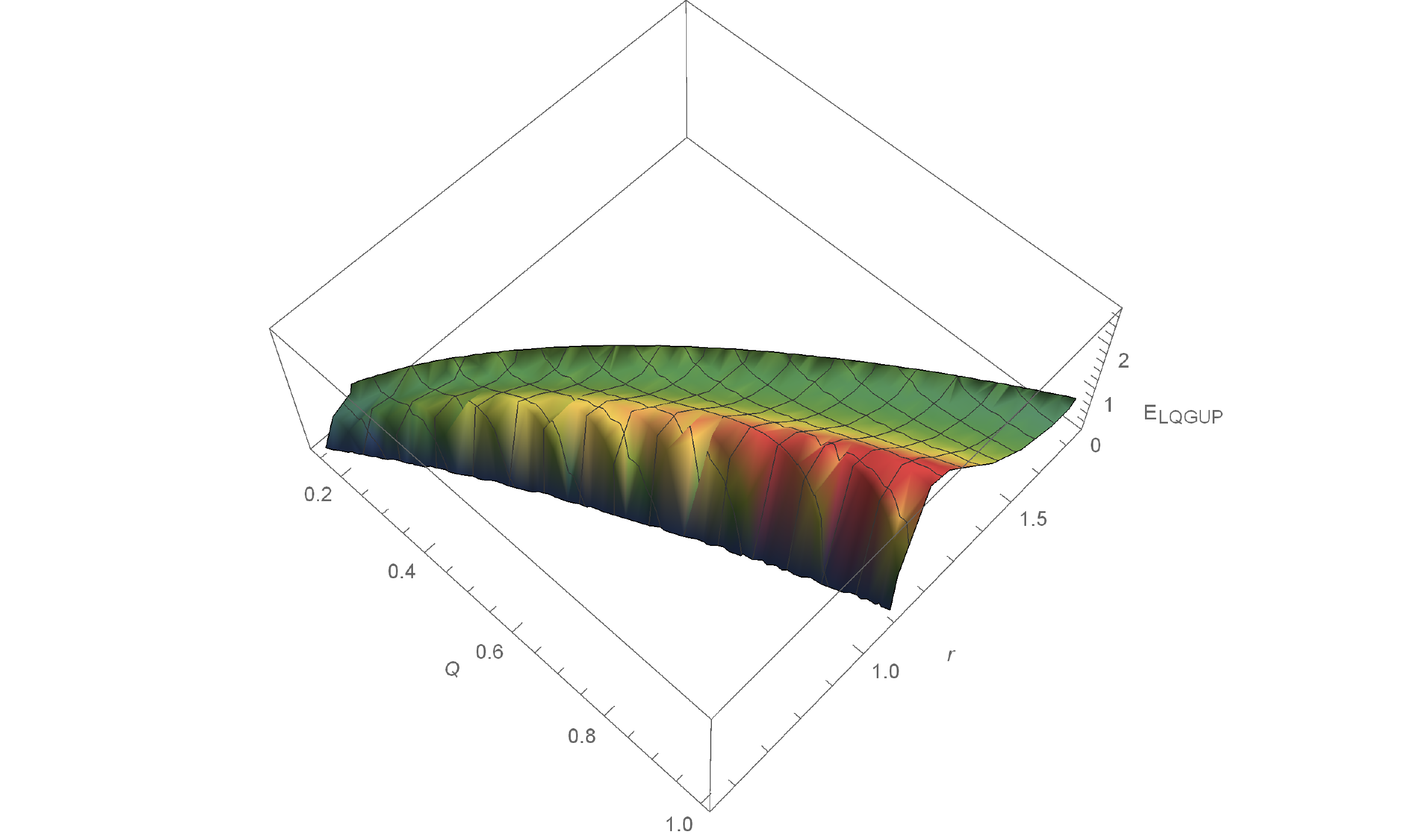}
\subcaption{ $\varpi=s=\frac{1}{2}, \Lambda=1$.}
\end{minipage}\hfill
\begin{minipage}[t]{0.45\linewidth}
\captionsetup{justification=centering}
\includegraphics[width=\linewidth]{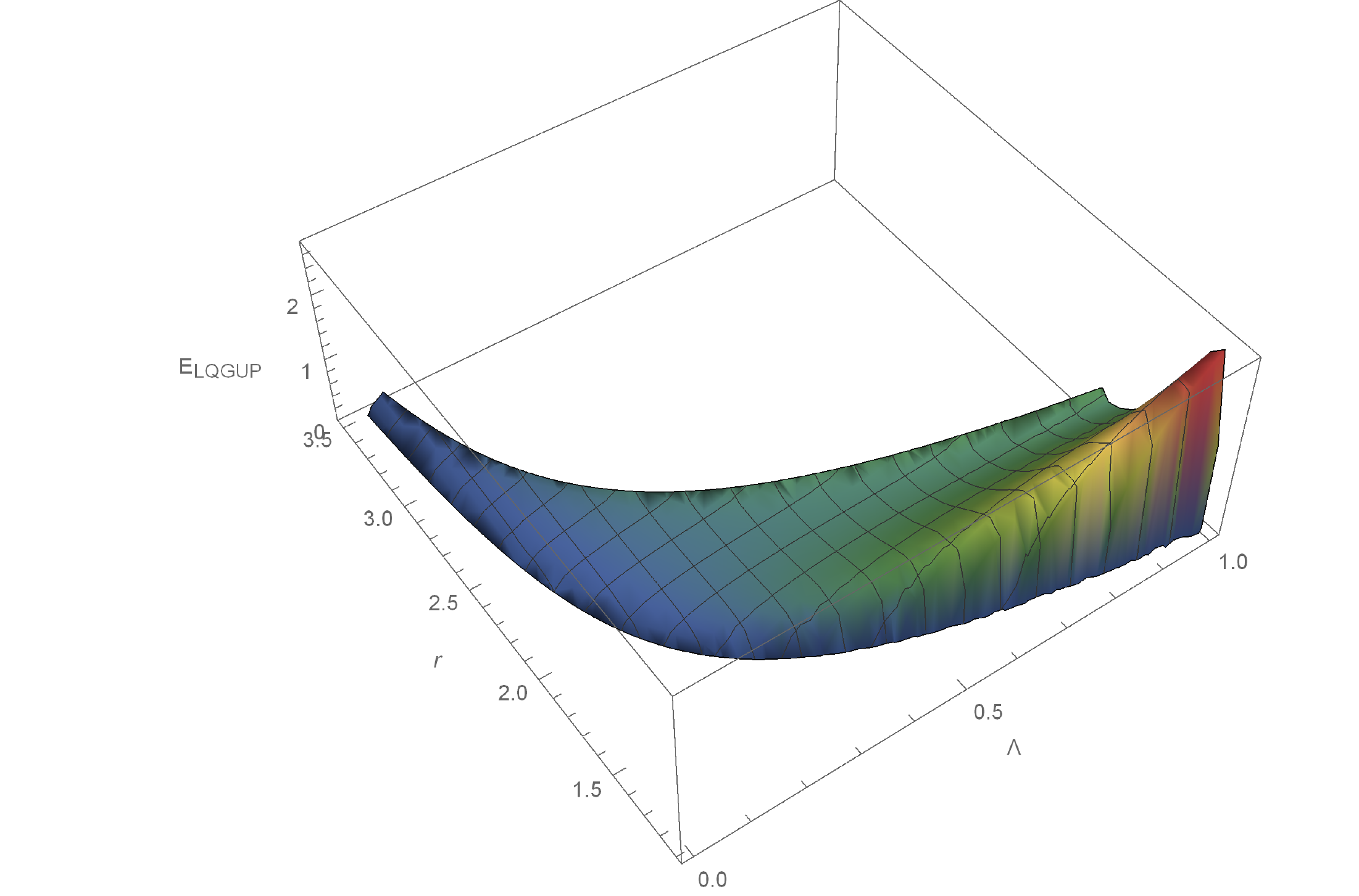}
\subcaption{ $\varpi=s=\frac{1}{2}, Q=1$.}
\end{minipage}

\begin{minipage}[t]{0.45\linewidth}
\captionsetup{justification=centering}
\includegraphics[width=\linewidth]{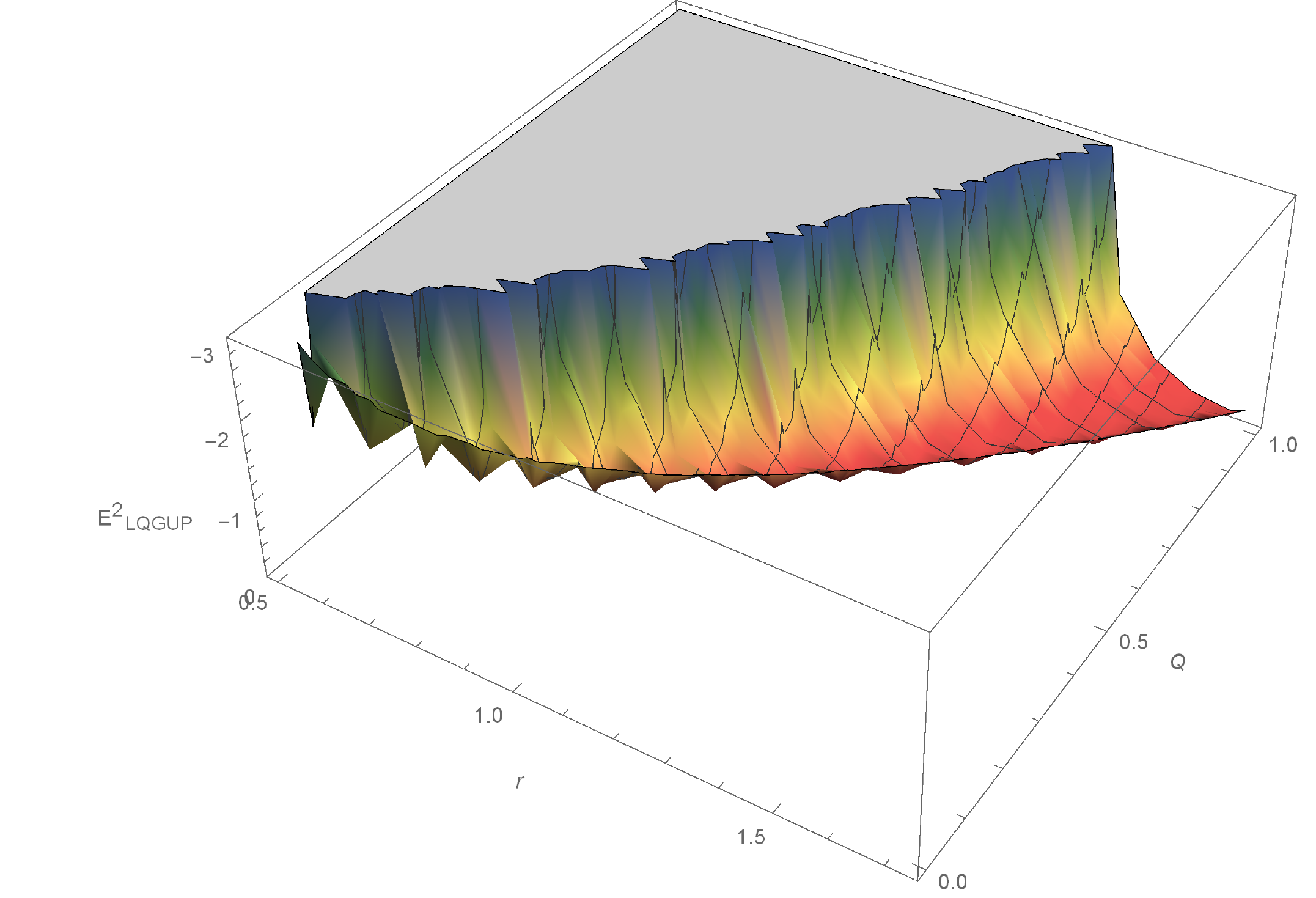}
\subcaption{ $\varpi=-(s=\frac{1}{2}), \Lambda=1$.}
\end{minipage}\hfill
\begin{minipage}[t]{0.45\linewidth}
\captionsetup{justification=centering}
\includegraphics[width=\linewidth]{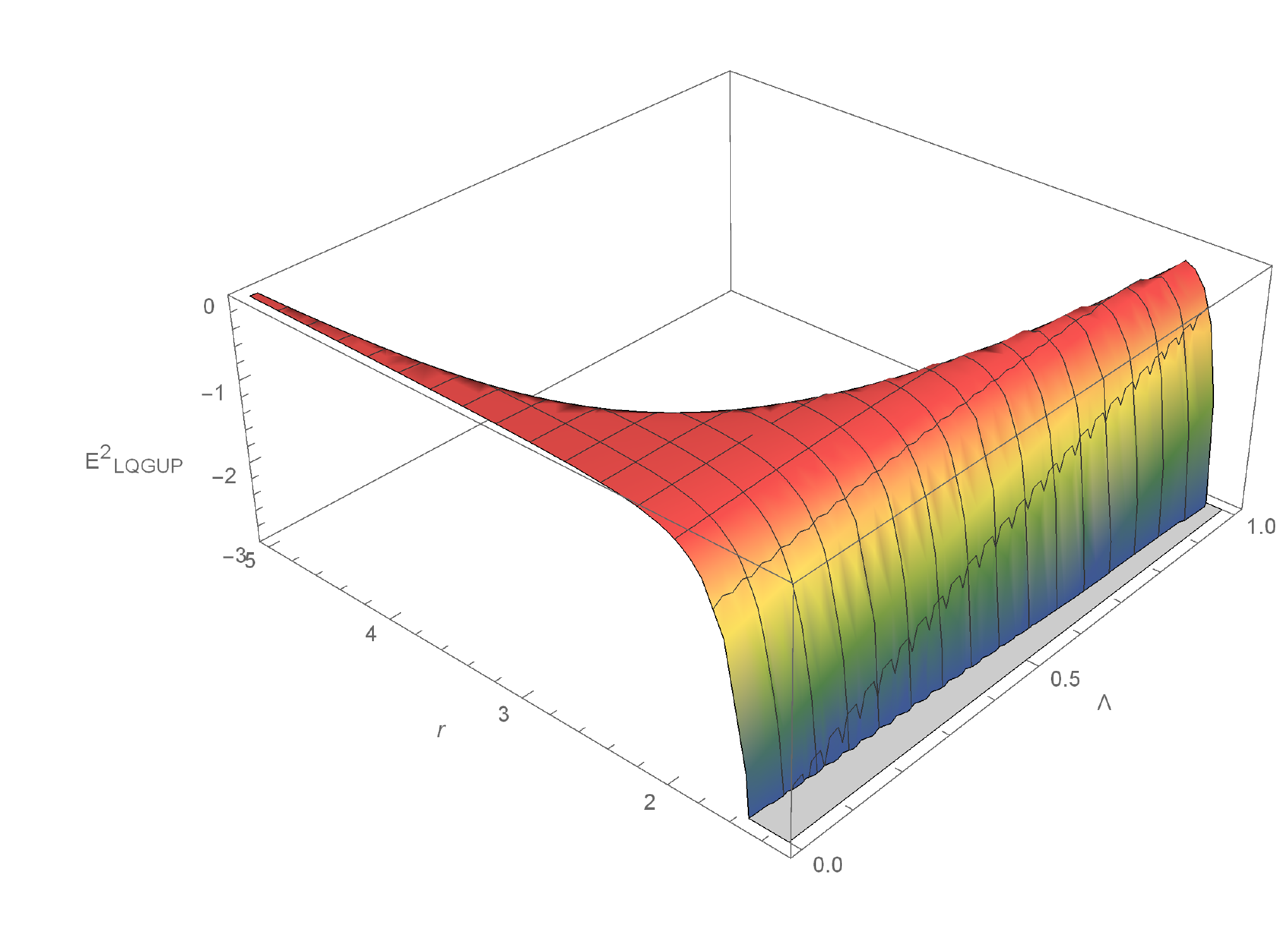}
\subcaption{ $\varpi=-(s=\frac{1}{2}), Q=1$.}
\end{minipage}

\begin{minipage}[t]{0.45\linewidth}
\captionsetup{justification=centering}
\includegraphics[width=\linewidth]{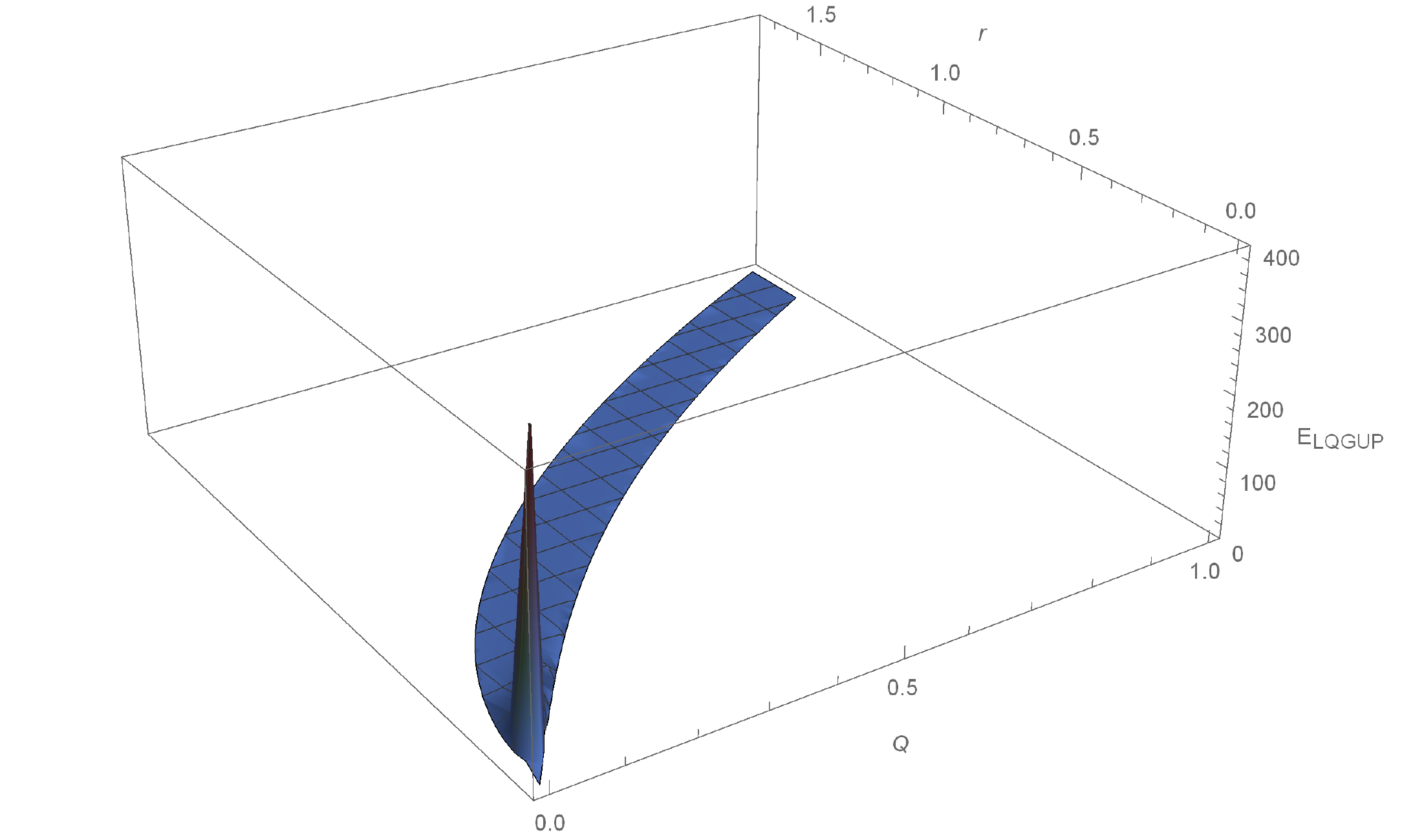}
\subcaption{ $\varpi=s=-\frac{1}{2}, \Lambda=1$.}
\end{minipage}\hfill
\begin{minipage}[t]{0.45\linewidth}
\captionsetup{justification=centering}
\includegraphics[width=\linewidth]{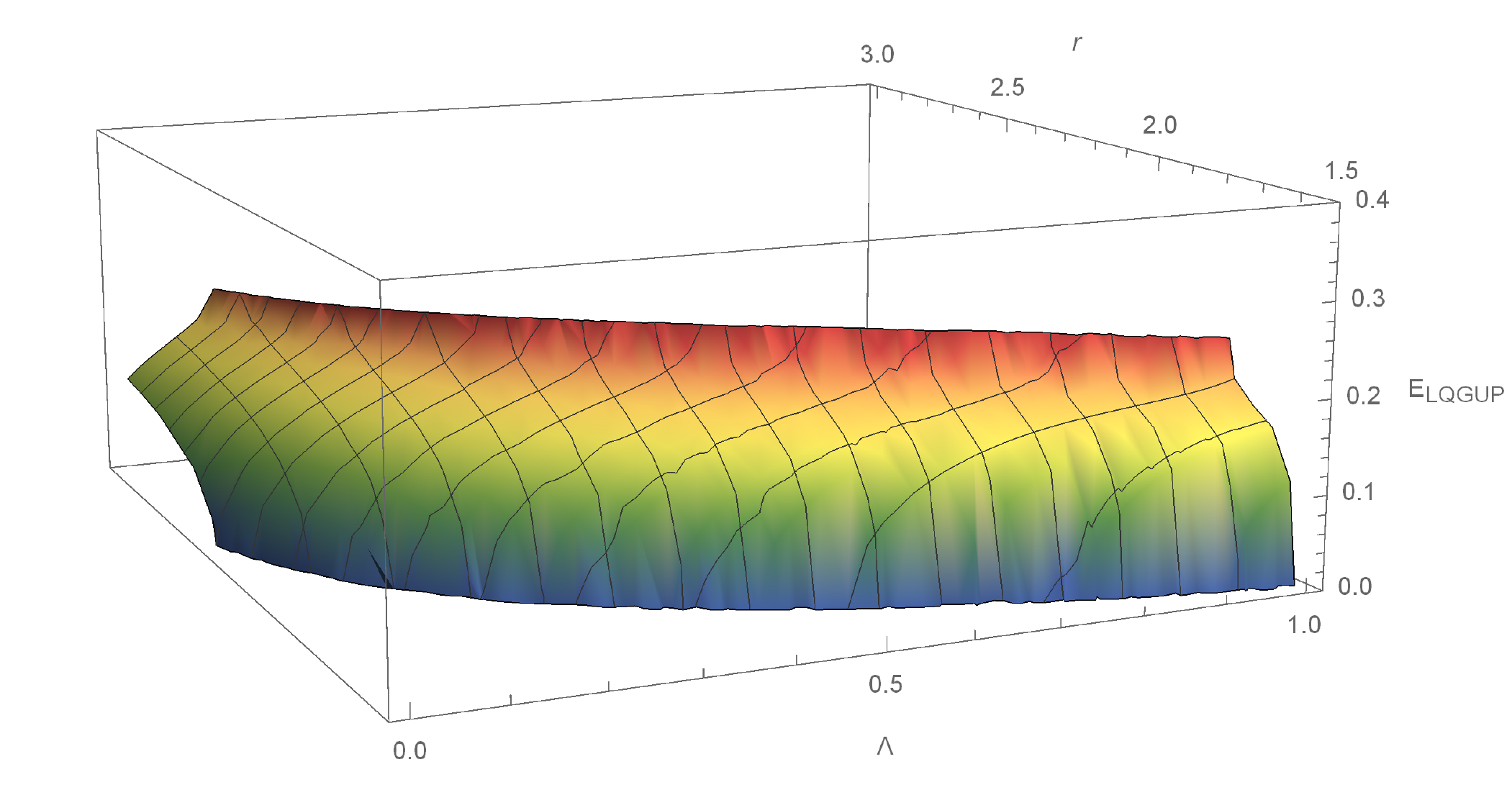}
\subcaption{ $\varpi=s=-\frac{1}{2}, Q=1$.}
\end{minipage}

\begin{minipage}[t]{0.45\linewidth}
\captionsetup{justification=centering}
\includegraphics[width=\linewidth]{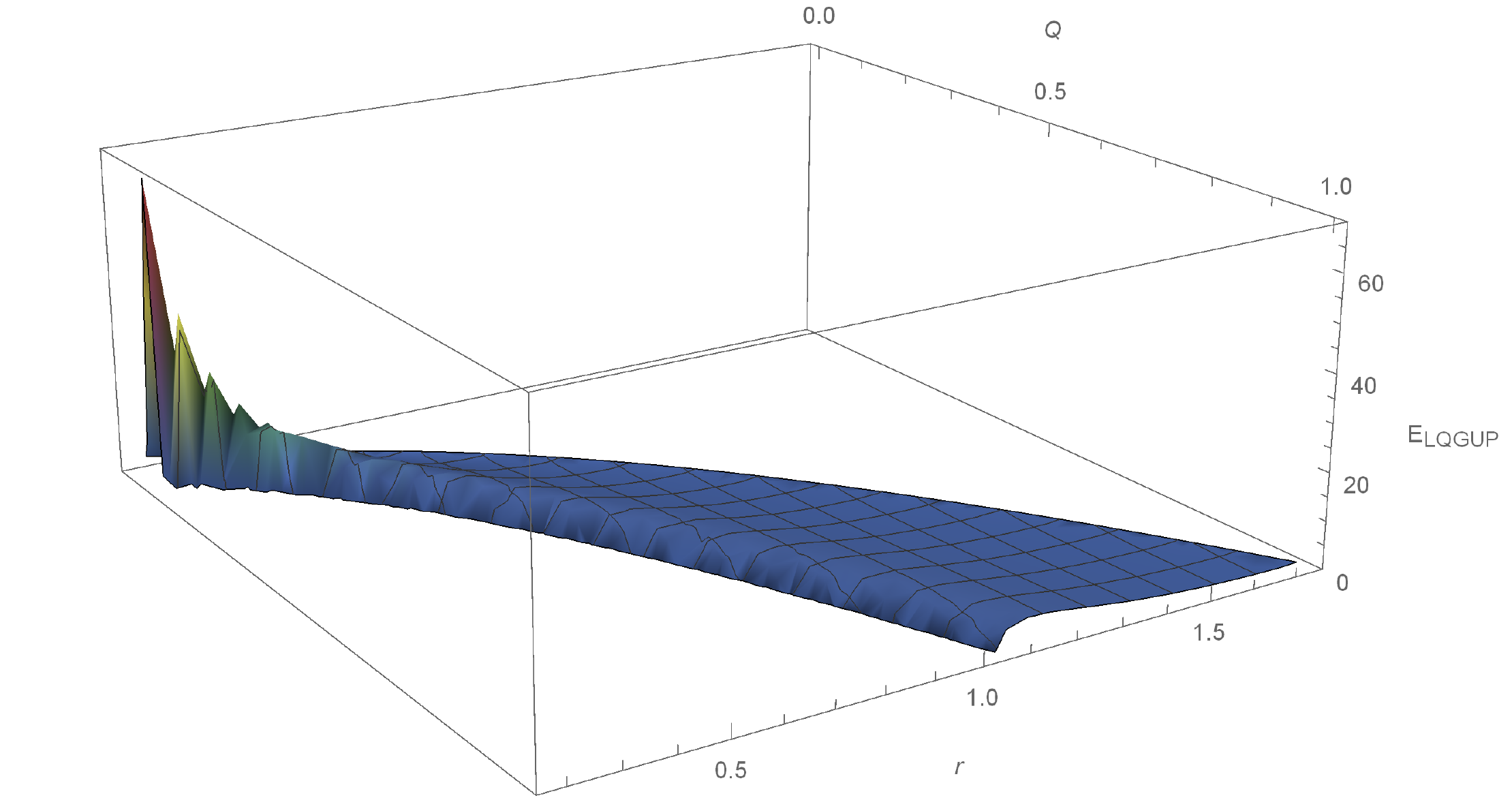}
\subcaption{ $\varpi=-(s=-\frac{1}{2}), \Lambda=1$.}
\end{minipage}\hfill
\begin{minipage}[t]{0.45\linewidth}
\captionsetup{justification=centering}
\includegraphics[width=\linewidth]{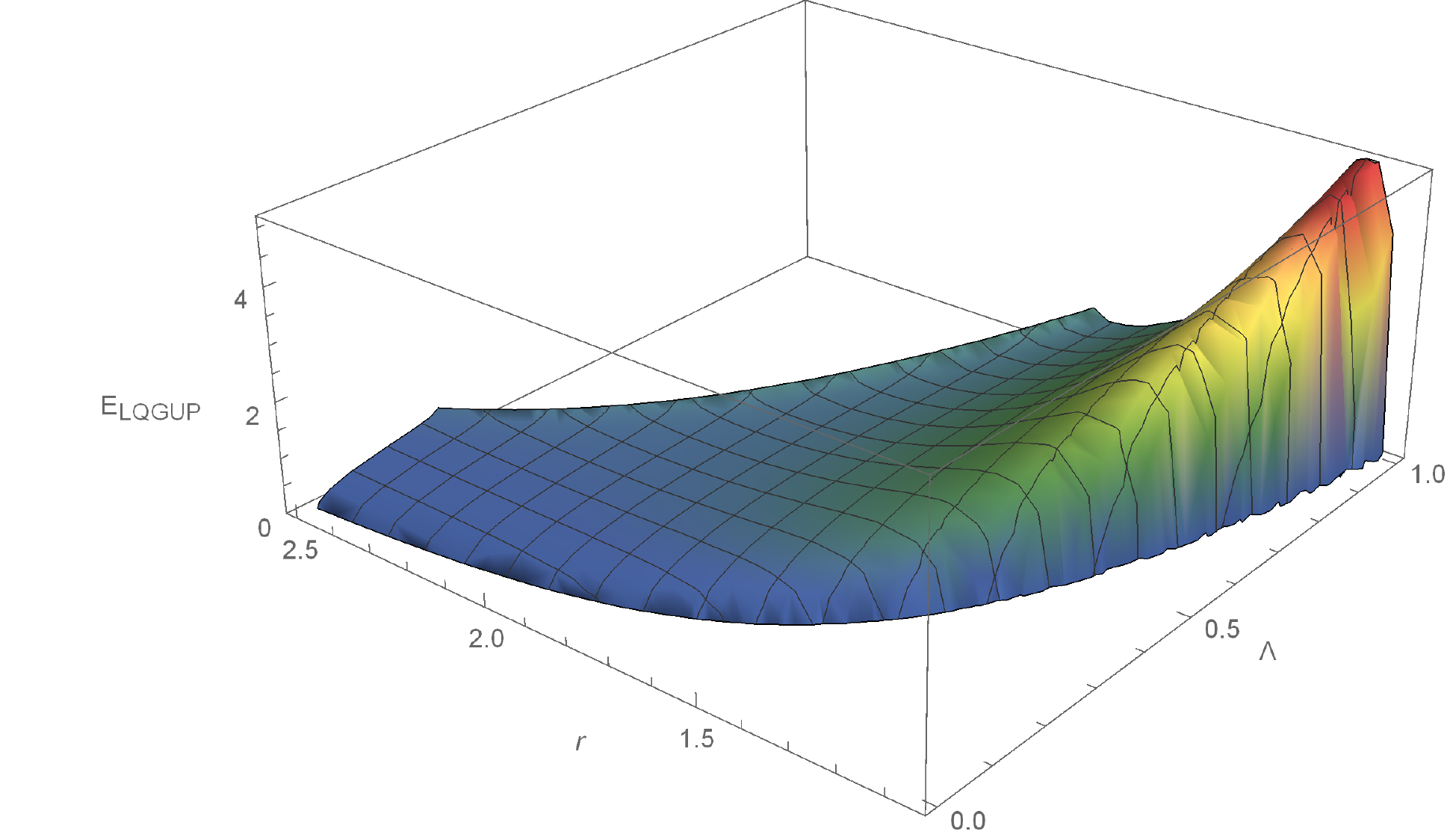}
\subcaption{ $\varpi=-(s=-\frac{1}{2}), Q=1$.}
\end{minipage}

\captionsetup{justification=centering}
\caption{Tunneling energy $E_{\text{LQGUP}}$ for $\varpi=\pm (s=\pm\frac{1}{2})$ fermionic massless fields.}\label{fig.2}
\end{figure}



\begin{figure}[h!]
\begin{minipage}[t]{0.45\linewidth}
\captionsetup{justification=centering}
\includegraphics[width=\linewidth]{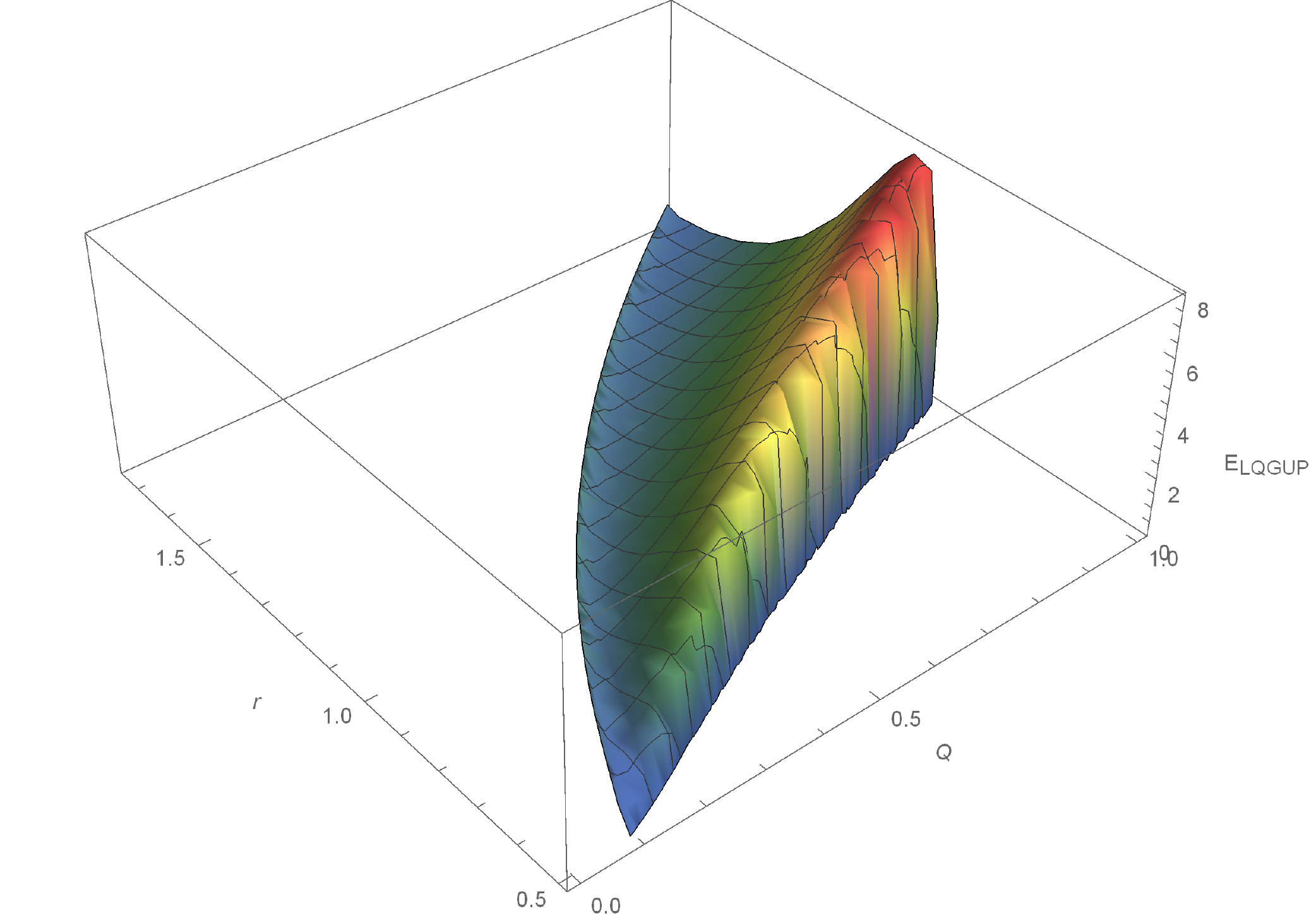}
\subcaption{ $\varpi=s=1, \Lambda=1$.}
\end{minipage}\hfill
\begin{minipage}[t]{0.45\linewidth}
\captionsetup{justification=centering}
\includegraphics[width=\linewidth]{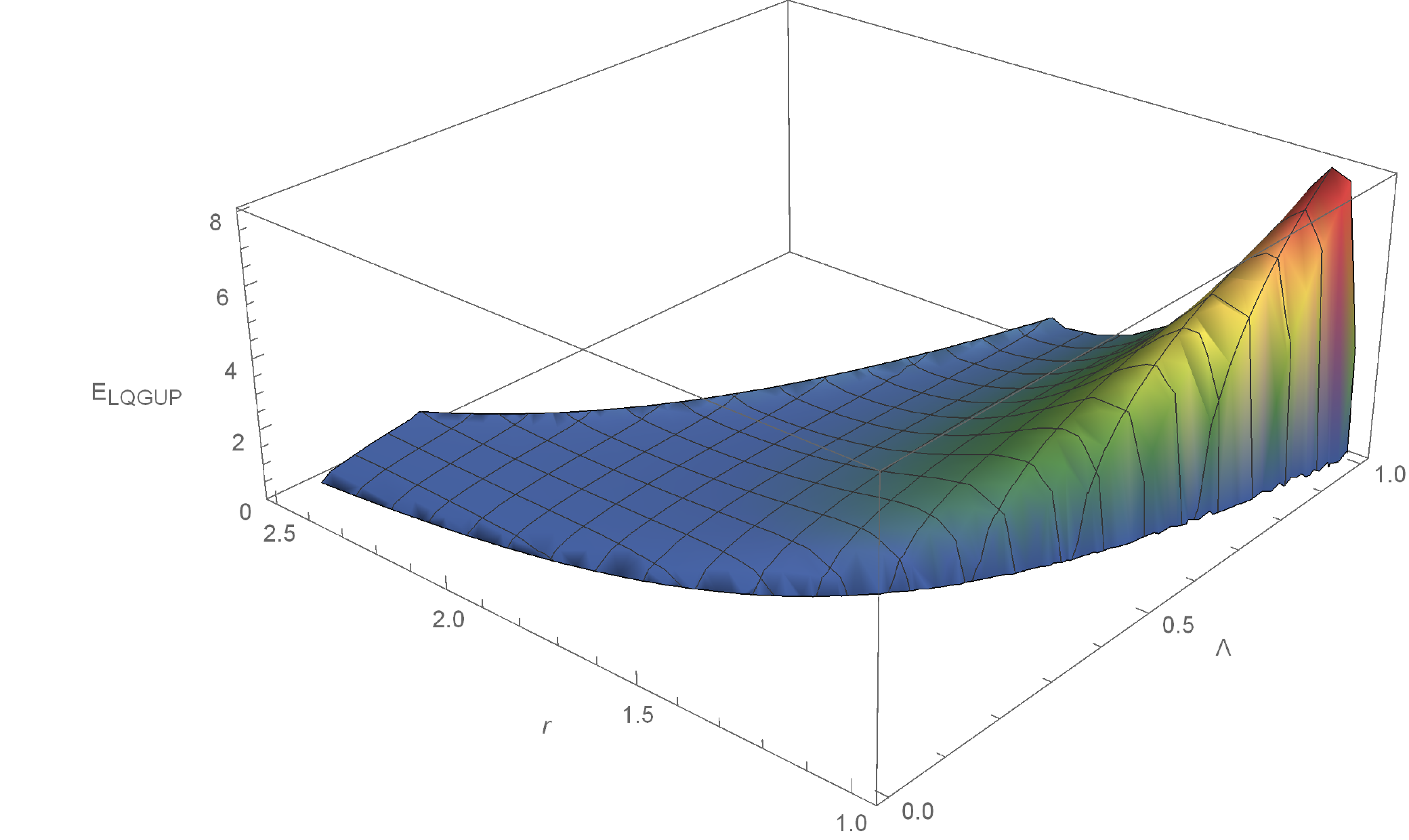}
\subcaption{ $\varpi=s=1, Q=1$.}
\end{minipage}

\begin{minipage}[t]{0.45\linewidth}
\captionsetup{justification=centering}
\includegraphics[width=\linewidth]{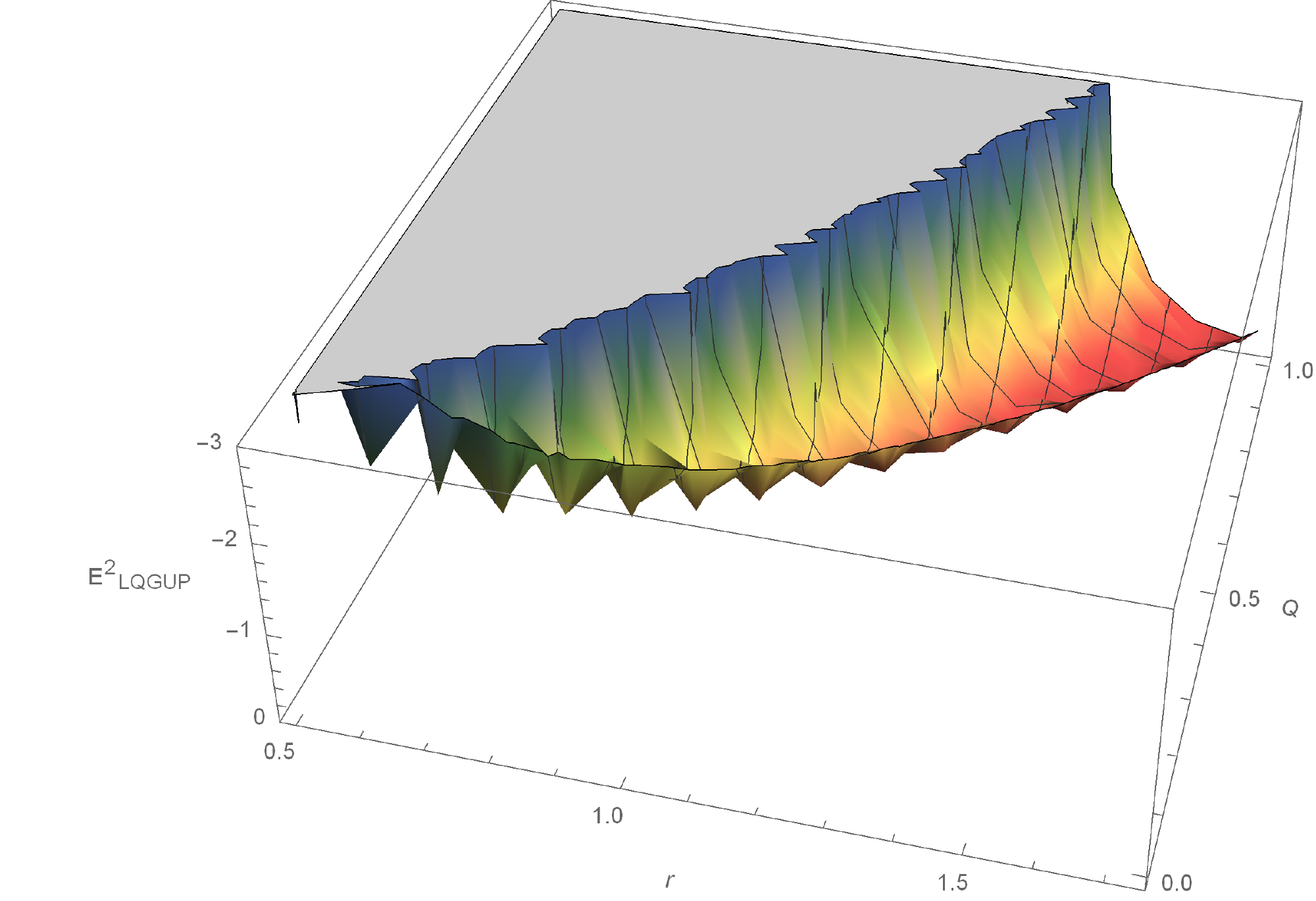}
\subcaption{ $\varpi=-(s=1), \Lambda=1$.}
\end{minipage}\hfill
\begin{minipage}[t]{0.45\linewidth}
\captionsetup{justification=centering}
\includegraphics[width=\linewidth]{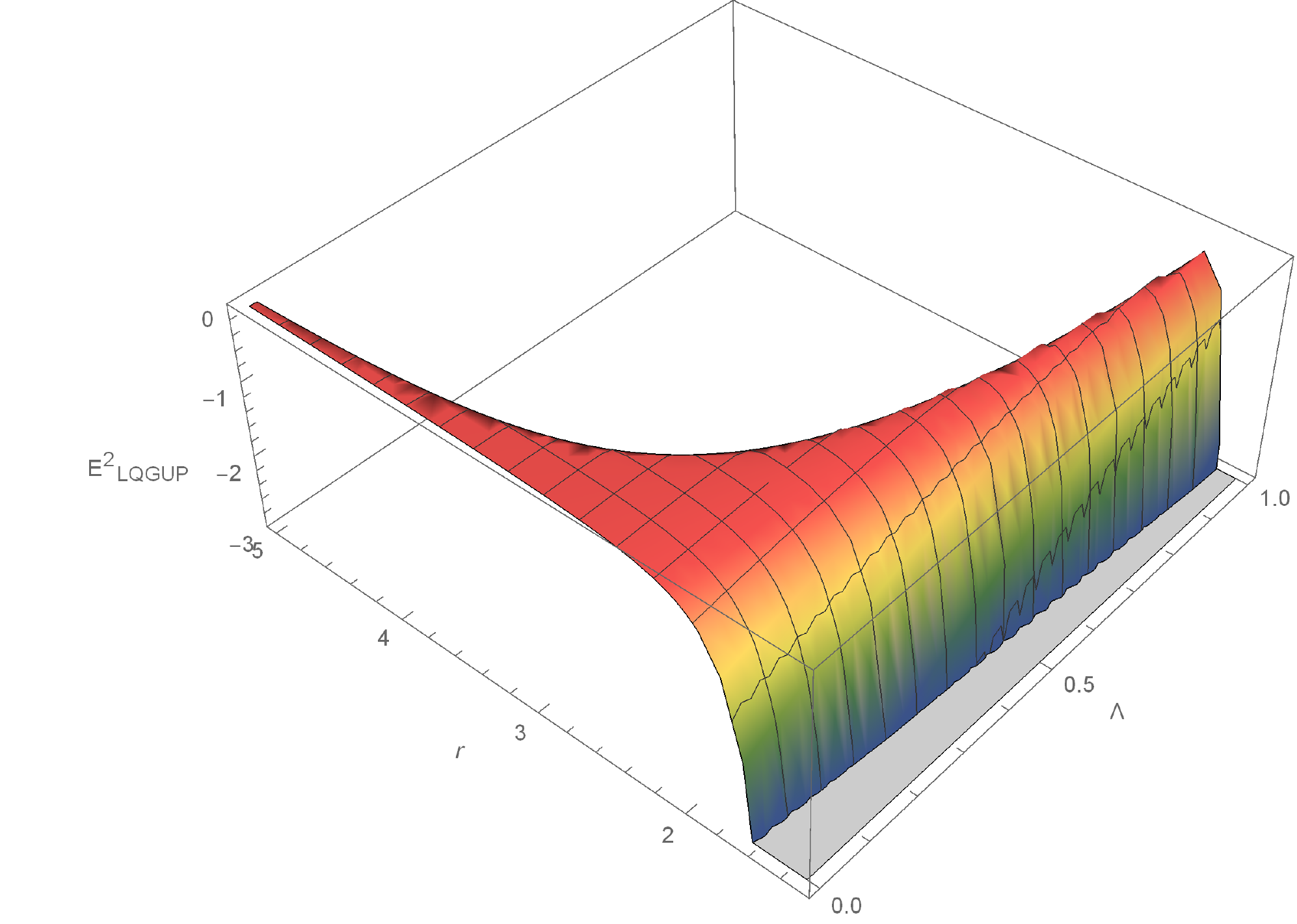}
\subcaption{ $\varpi=-(s=1), Q=1$.}
\end{minipage}

\begin{minipage}[t]{0.45\linewidth}
\captionsetup{justification=centering}
\includegraphics[width=\linewidth]{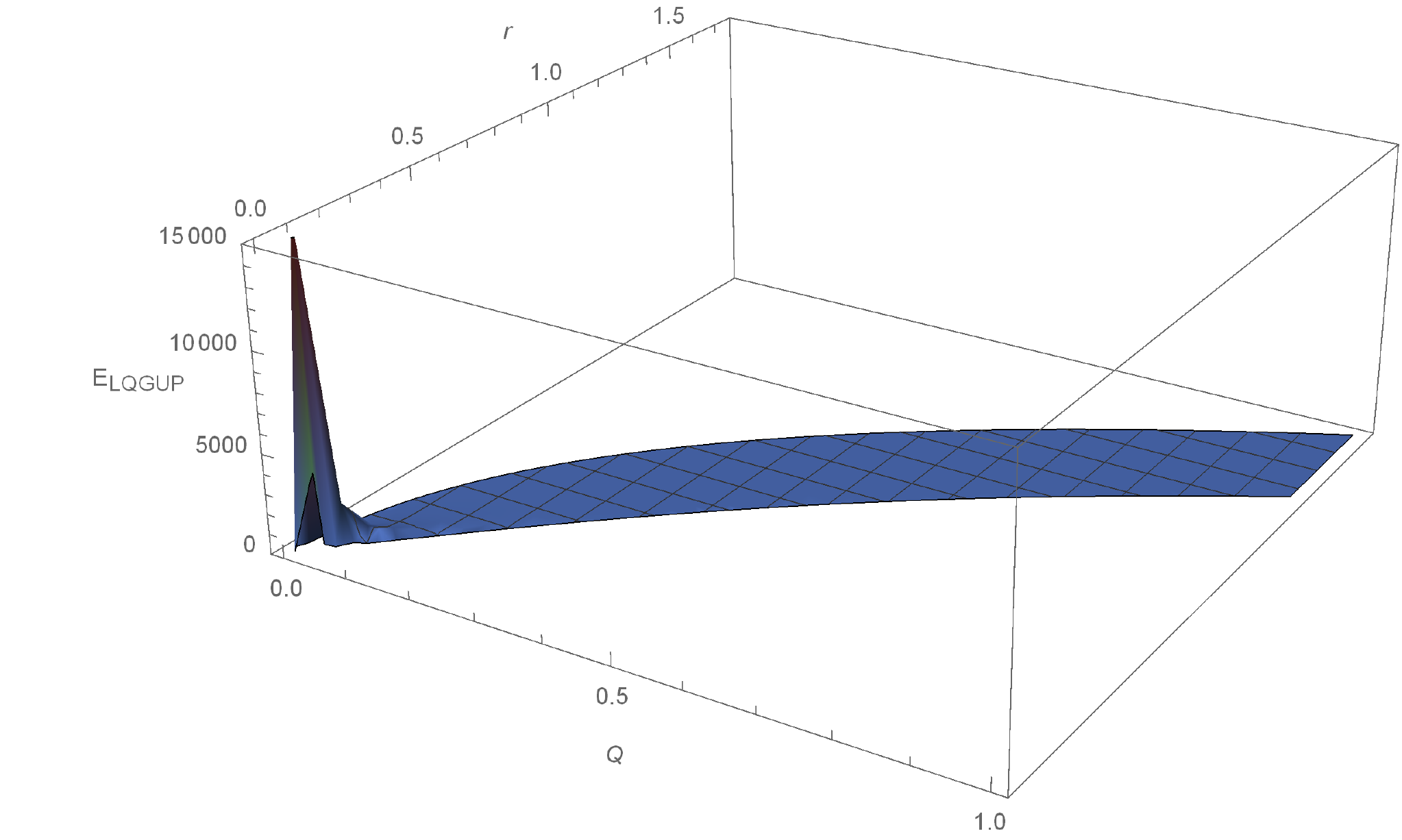}
\subcaption{ $\varpi=s=-1, \Lambda=1$.}
\end{minipage}\hfill
\begin{minipage}[t]{0.45\linewidth}
\captionsetup{justification=centering}
\includegraphics[width=\linewidth]{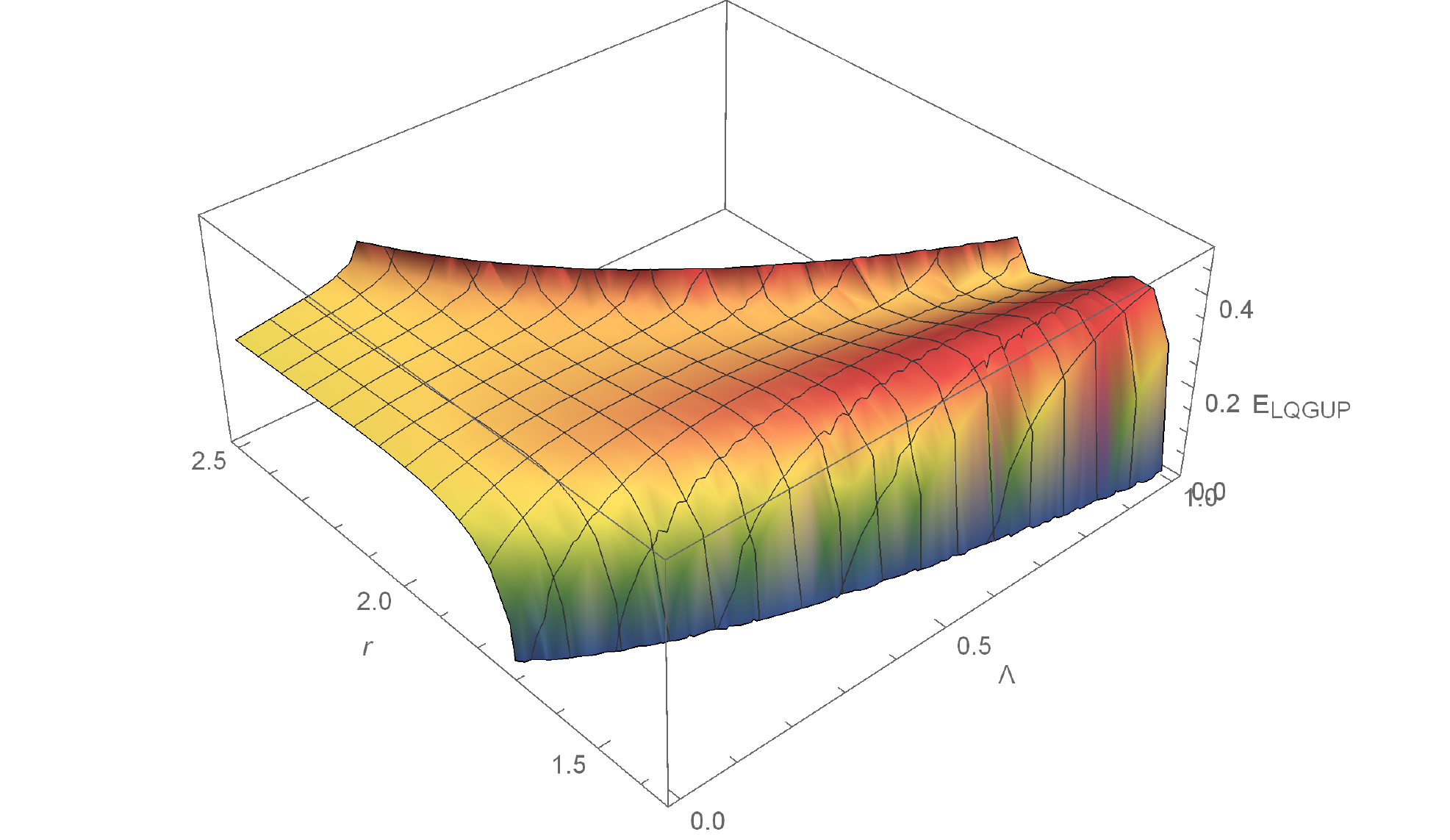}
\subcaption{ $\varpi=s=-1, Q=1$.}
\end{minipage}

\begin{minipage}[t]{0.45\linewidth}
\captionsetup{justification=centering}
\includegraphics[width=\linewidth]{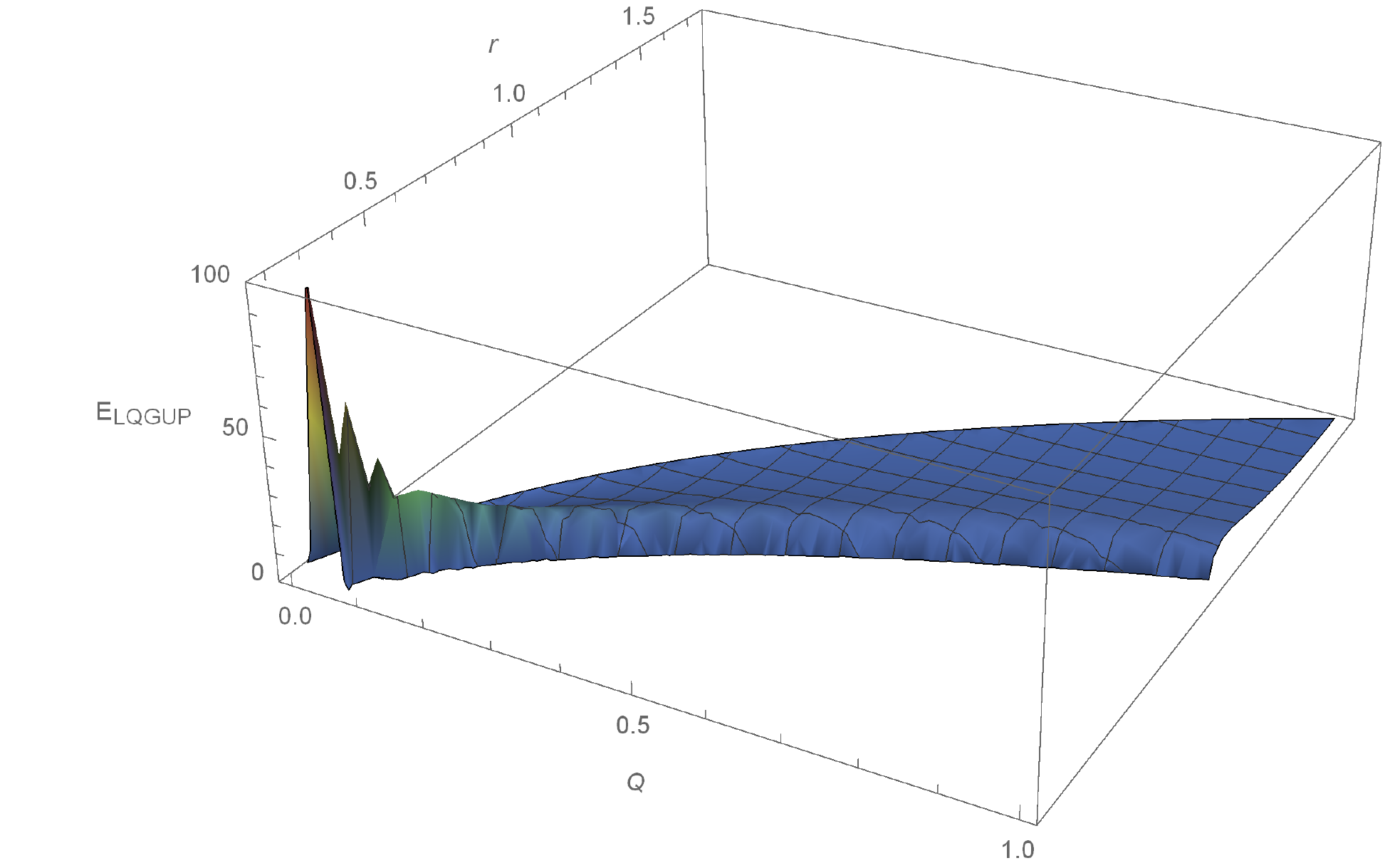}
\subcaption{ $\varpi=-(s=-1), \Lambda=1$.}
\end{minipage}\hfill
\begin{minipage}[t]{0.45\linewidth}
\captionsetup{justification=centering}
\includegraphics[width=\linewidth]{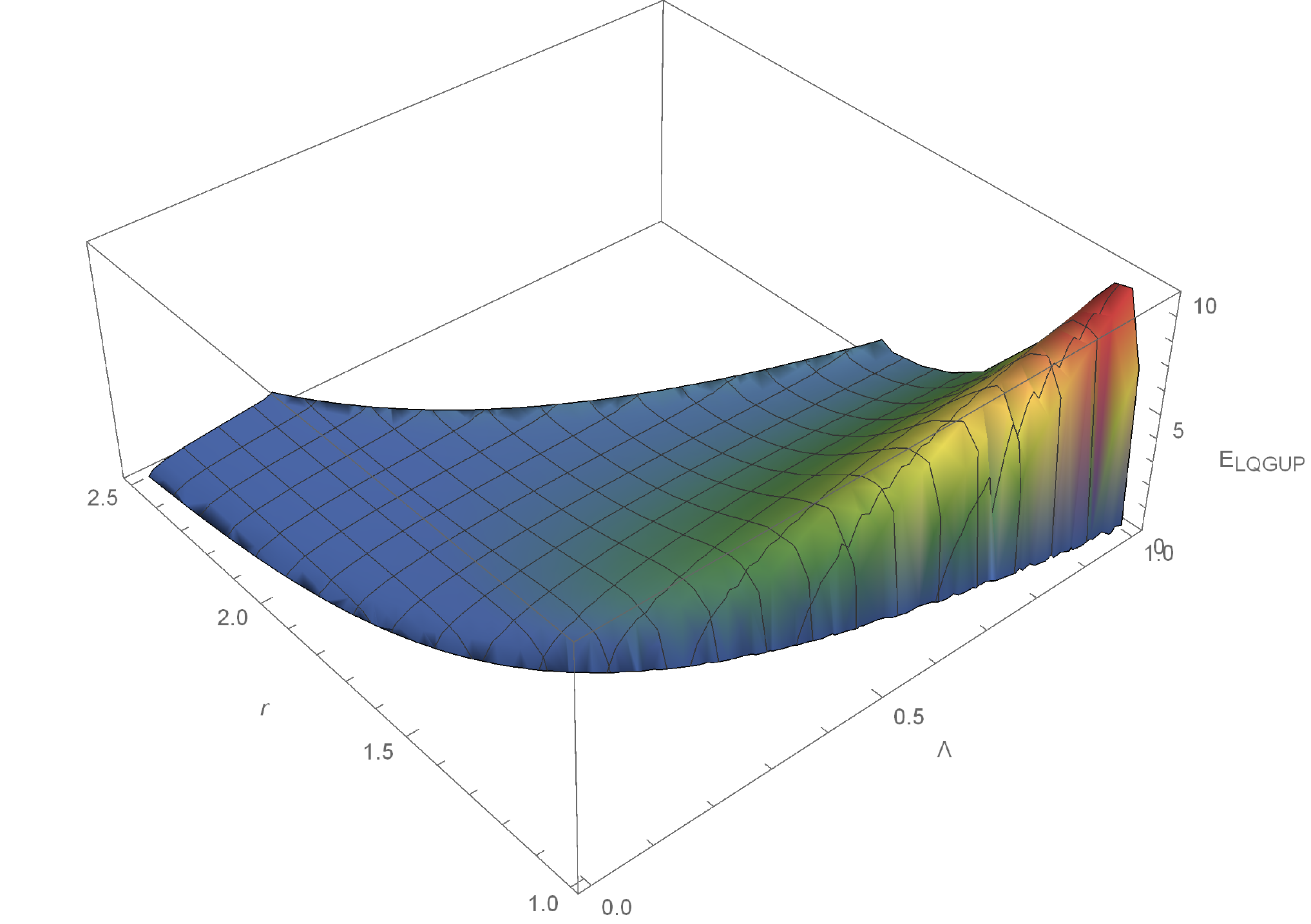}
\subcaption{ $\varpi=-(s=-1), Q=1$.}
\end{minipage}

\captionsetup{justification=centering}
\caption{Tunneling energy $E_{\text{LQGUP}}$ for $\varpi=\pm (s=\pm 1)$ bosonic massless fields.}\label{fig.3}
\end{figure}


\begin{figure}[h!]
\begin{minipage}[t]{0.45\linewidth}
\captionsetup{justification=centering}
\includegraphics[width=\linewidth]{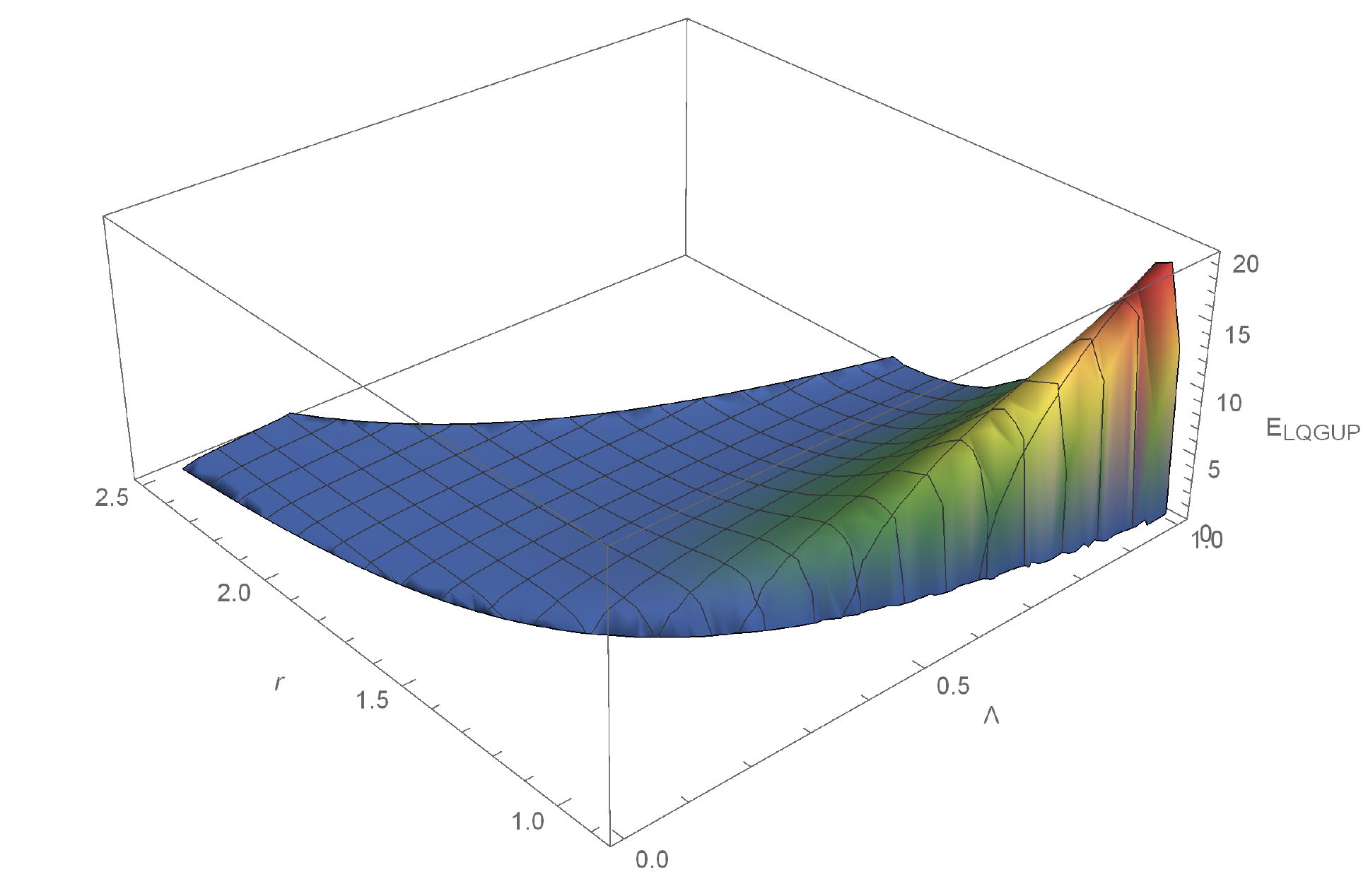}
\subcaption{ $\varpi=s=\frac{3}{2}, \Lambda=1$.}
\end{minipage}\hfill
\begin{minipage}[t]{0.45\linewidth}
\captionsetup{justification=centering}
\includegraphics[width=\linewidth]{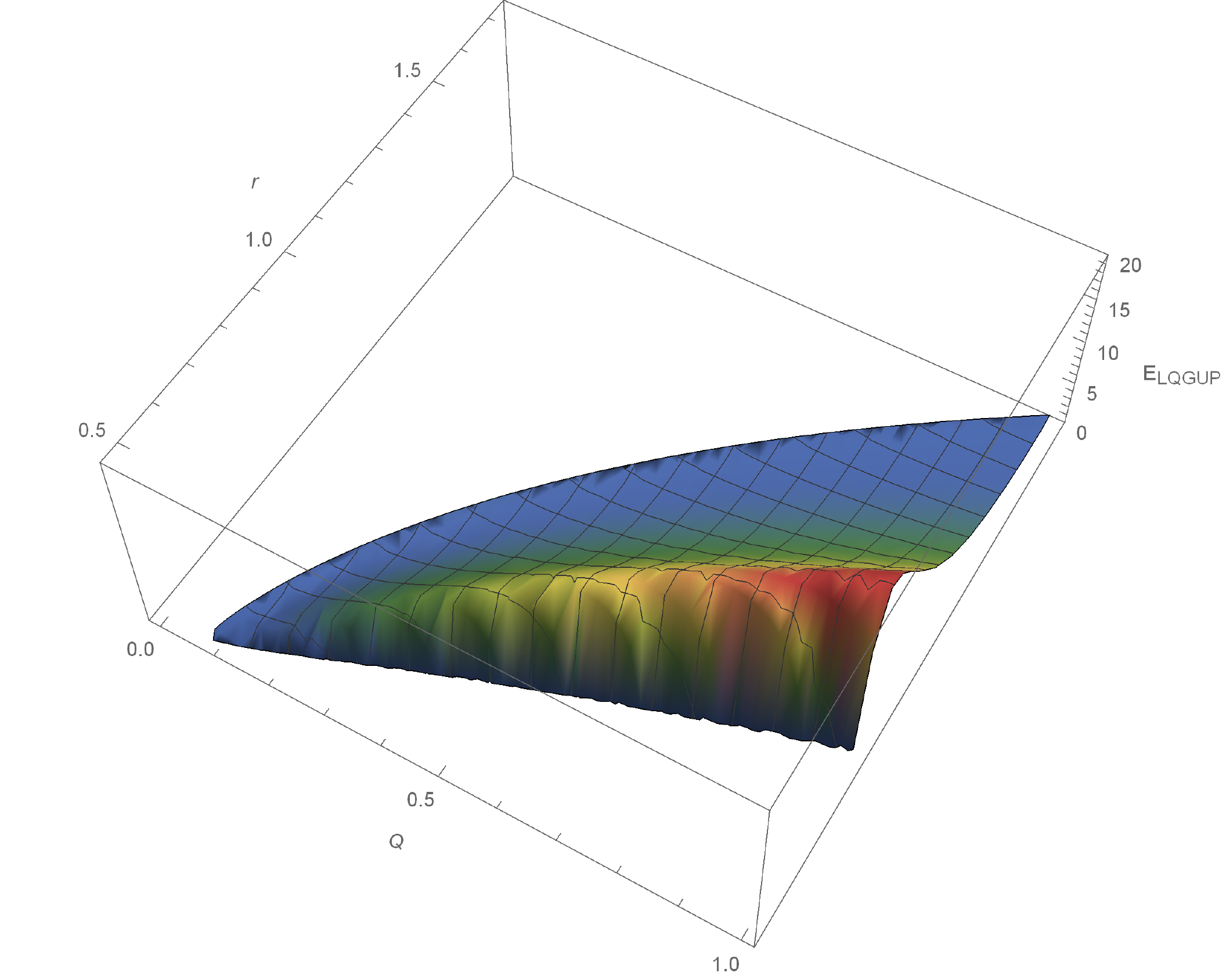}
\subcaption{ $\varpi=s=\frac{3}{2}, Q=1$.}
\end{minipage}

\begin{minipage}[t]{0.45\linewidth}
\captionsetup{justification=centering}
\includegraphics[width=\linewidth]{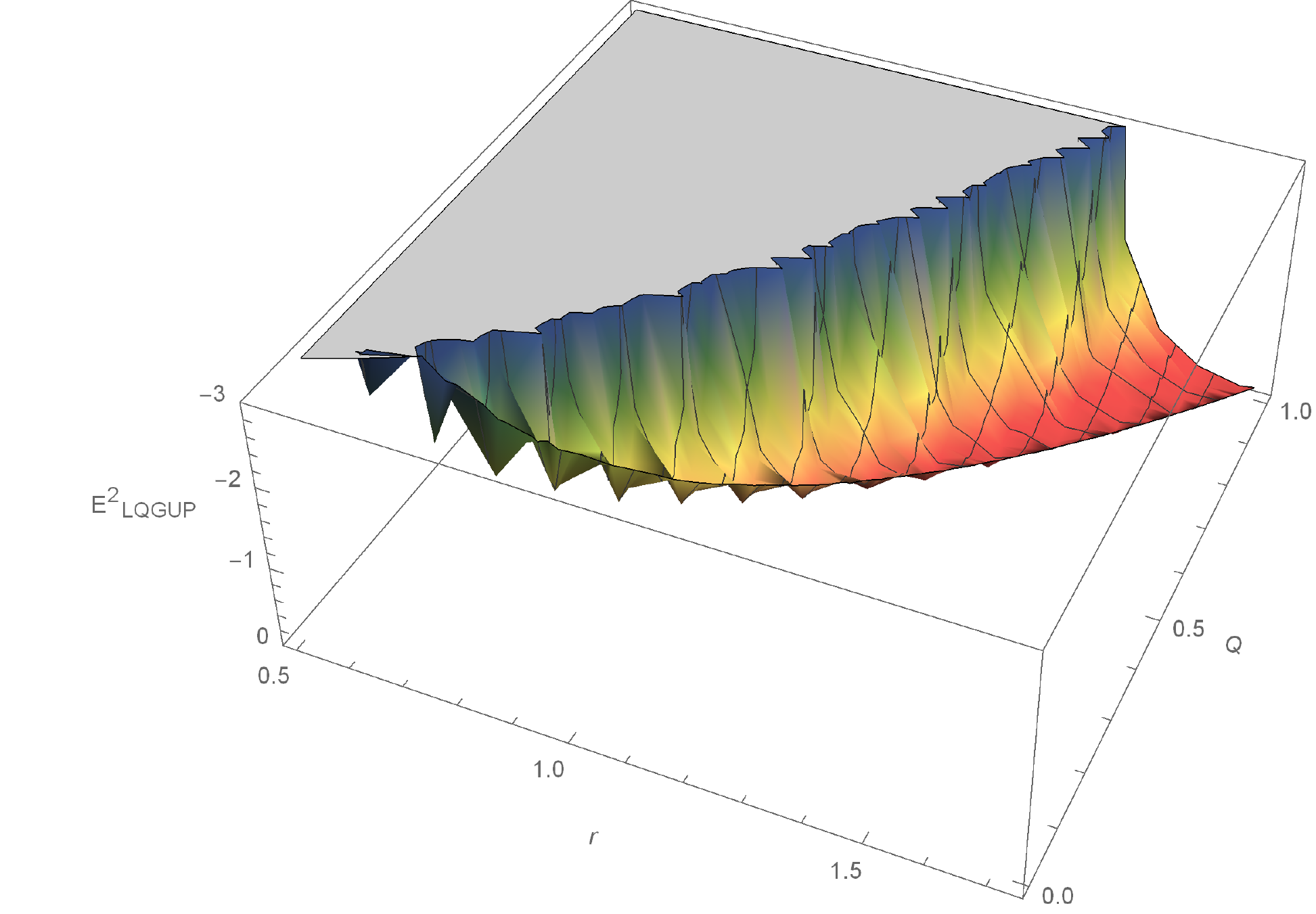}
\subcaption{ $\varpi=-(s=\frac{3}{2}), \Lambda=1$.}
\end{minipage}\hfill
\begin{minipage}[t]{0.45\linewidth}
\captionsetup{justification=centering}
\includegraphics[width=\linewidth]{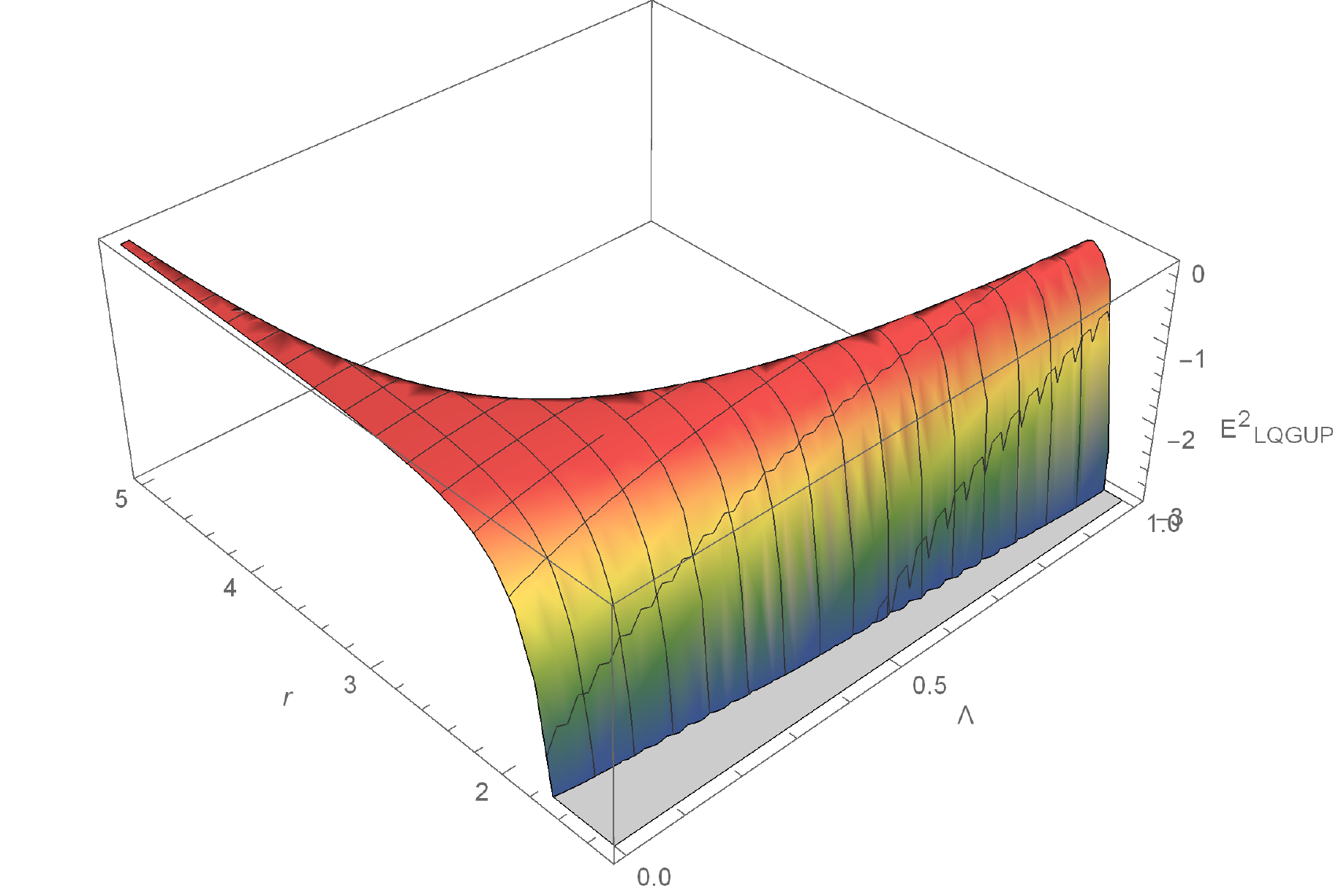}
\subcaption{ $\varpi=-(s=\frac{3}{2}), Q=1$.}
\end{minipage}

\begin{minipage}[t]{0.45\linewidth}
\captionsetup{justification=centering}
\includegraphics[width=\linewidth]{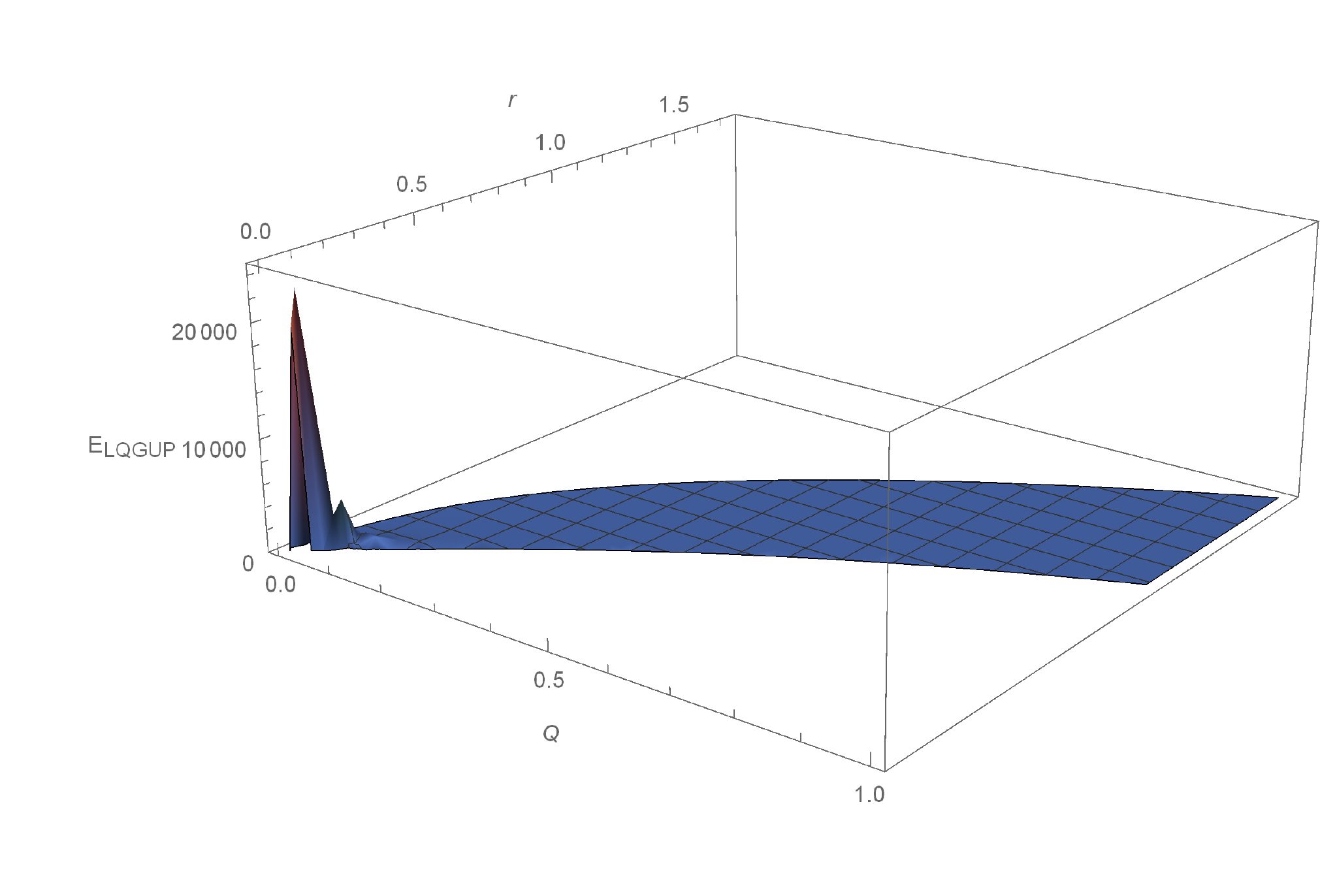}
\subcaption{ $\varpi=s=-\frac{3}{2}, \Lambda=1$.}
\end{minipage}\hfill
\begin{minipage}[t]{0.45\linewidth}
\captionsetup{justification=centering}
\includegraphics[width=\linewidth]{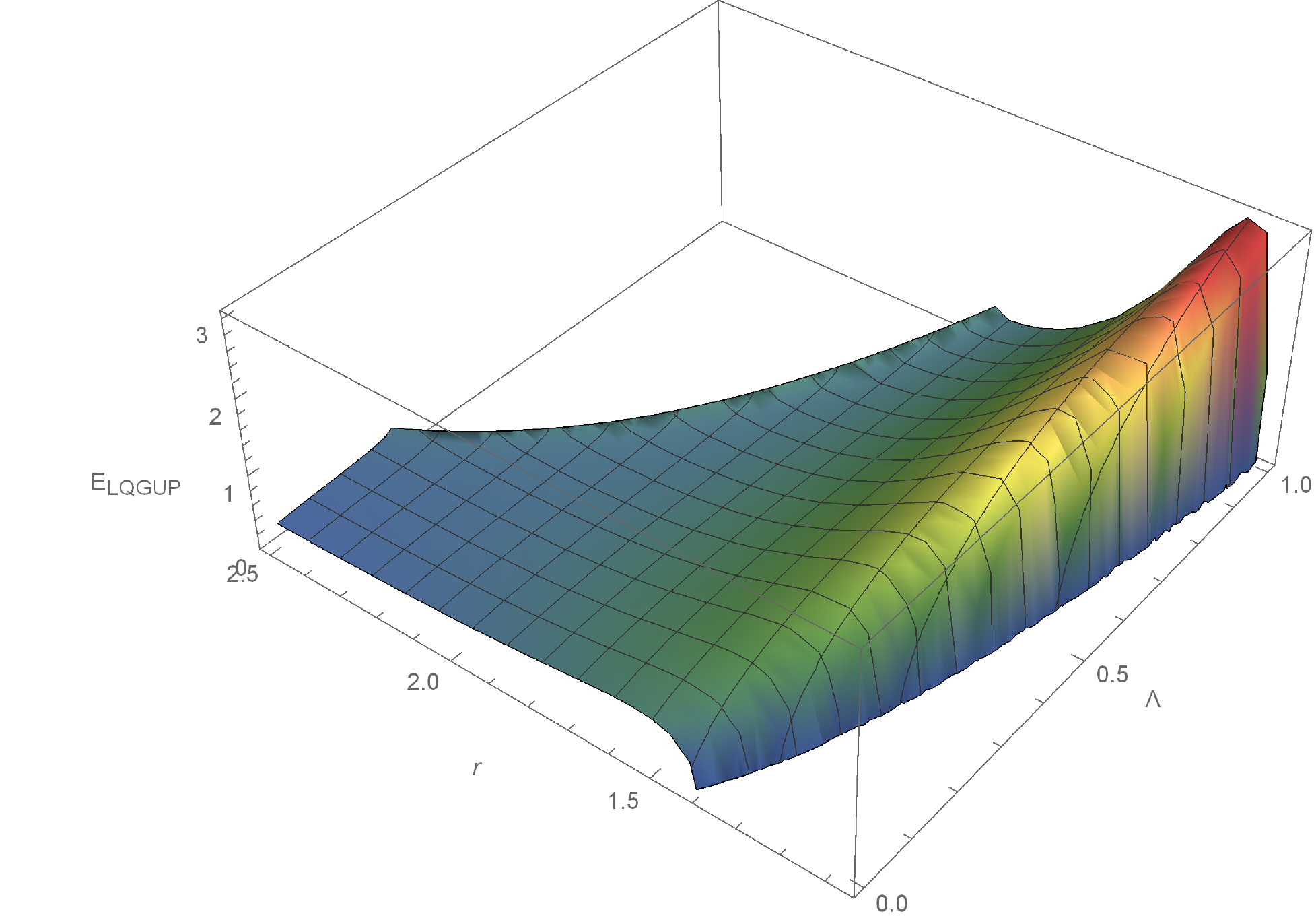}
\subcaption{ $\varpi=s=-\frac{3}{2}, Q=1$.}
\end{minipage}

\begin{minipage}[t]{0.45\linewidth}
\captionsetup{justification=centering}
\includegraphics[width=\linewidth]{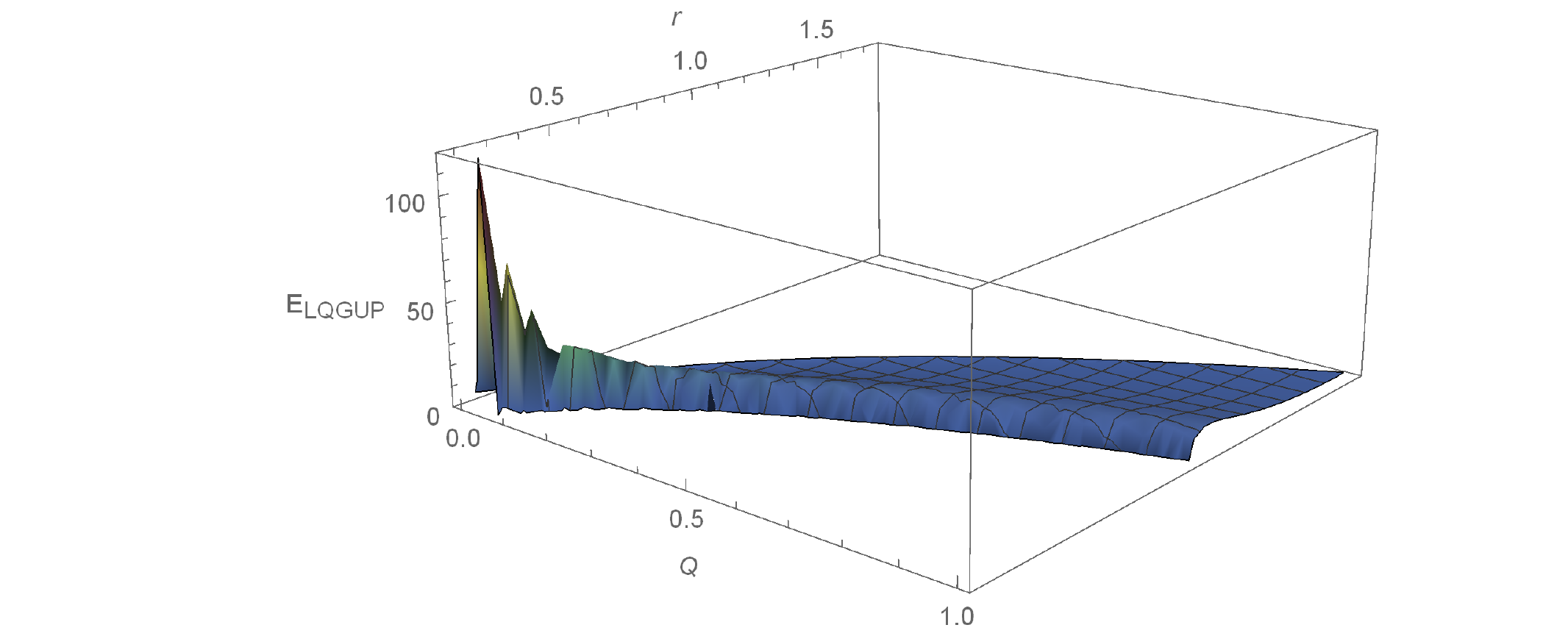}
\subcaption{ $\varpi=-(s=-\frac{3}{2}), \Lambda=1$.}
\end{minipage}\hfill
\begin{minipage}[t]{0.45\linewidth}
\captionsetup{justification=centering}
\includegraphics[width=\linewidth]{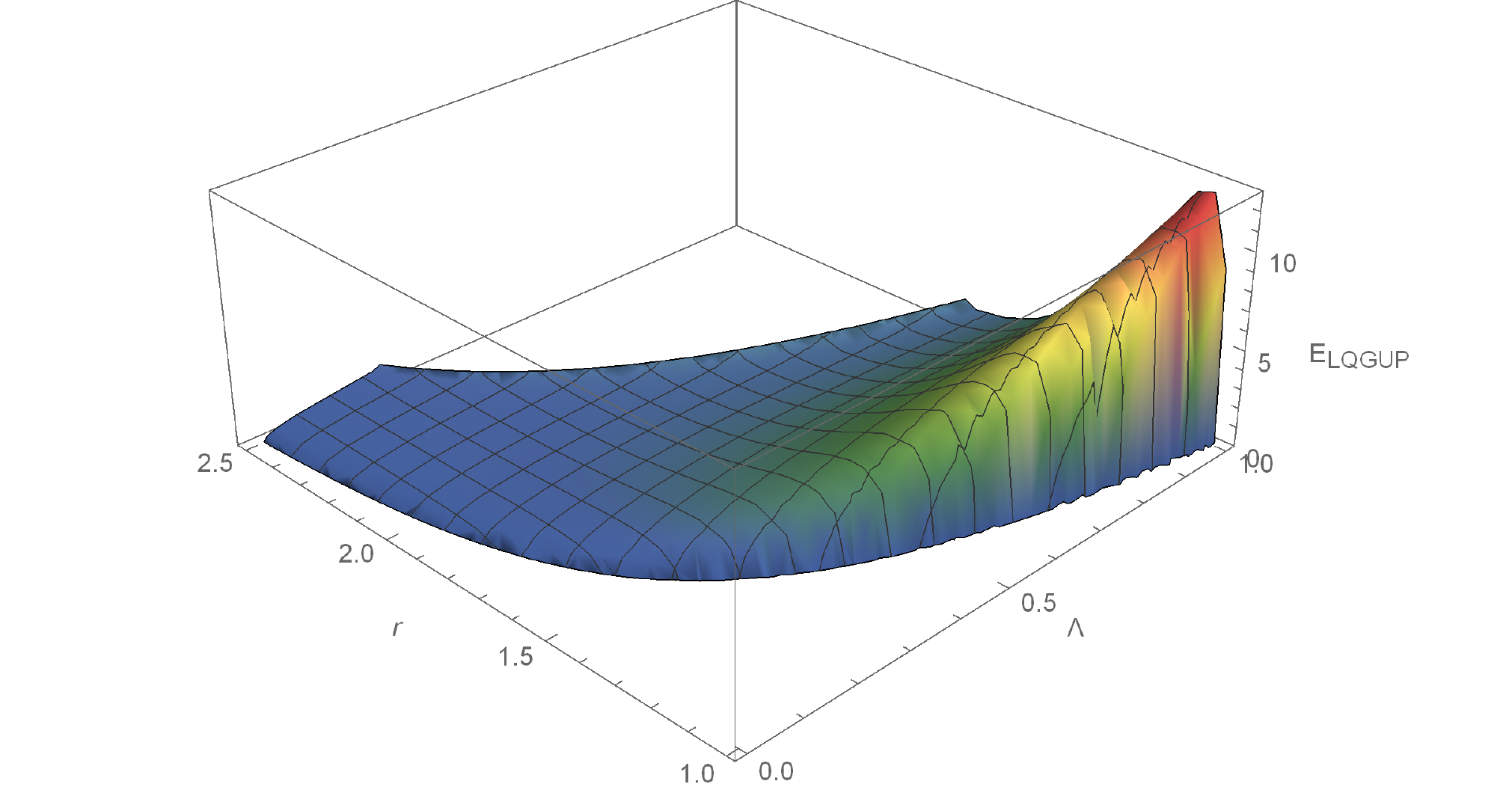}
\subcaption{ $\varpi=-(s=-\frac{3}{2}), Q=1$.}
\end{minipage}

\captionsetup{justification=centering}
\caption{Tunneling energy $E_{\text{LQGUP}}$ for $\varpi=\pm (s=\pm\frac{3}{2})$ fermionic massless fields.}\label{fig.4}
\end{figure}


\begin{figure}[h!]
\begin{minipage}[t]{0.45\linewidth}
\captionsetup{justification=centering}
\includegraphics[width=\linewidth]{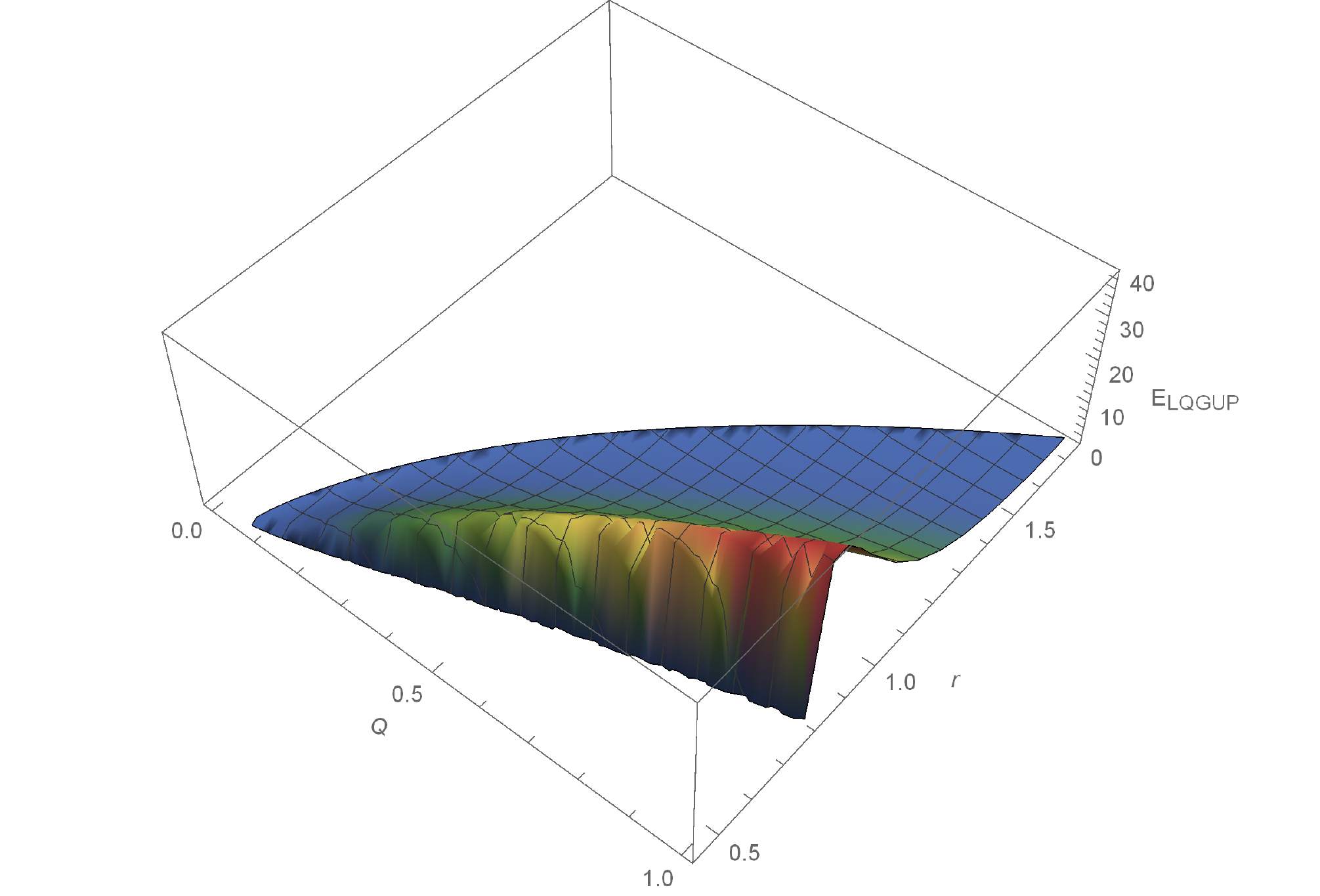}
\subcaption{ $\varpi=s=2, \Lambda=1$.}
\end{minipage}\hfill
\begin{minipage}[t]{0.45\linewidth}
\captionsetup{justification=centering}
\includegraphics[width=\linewidth]{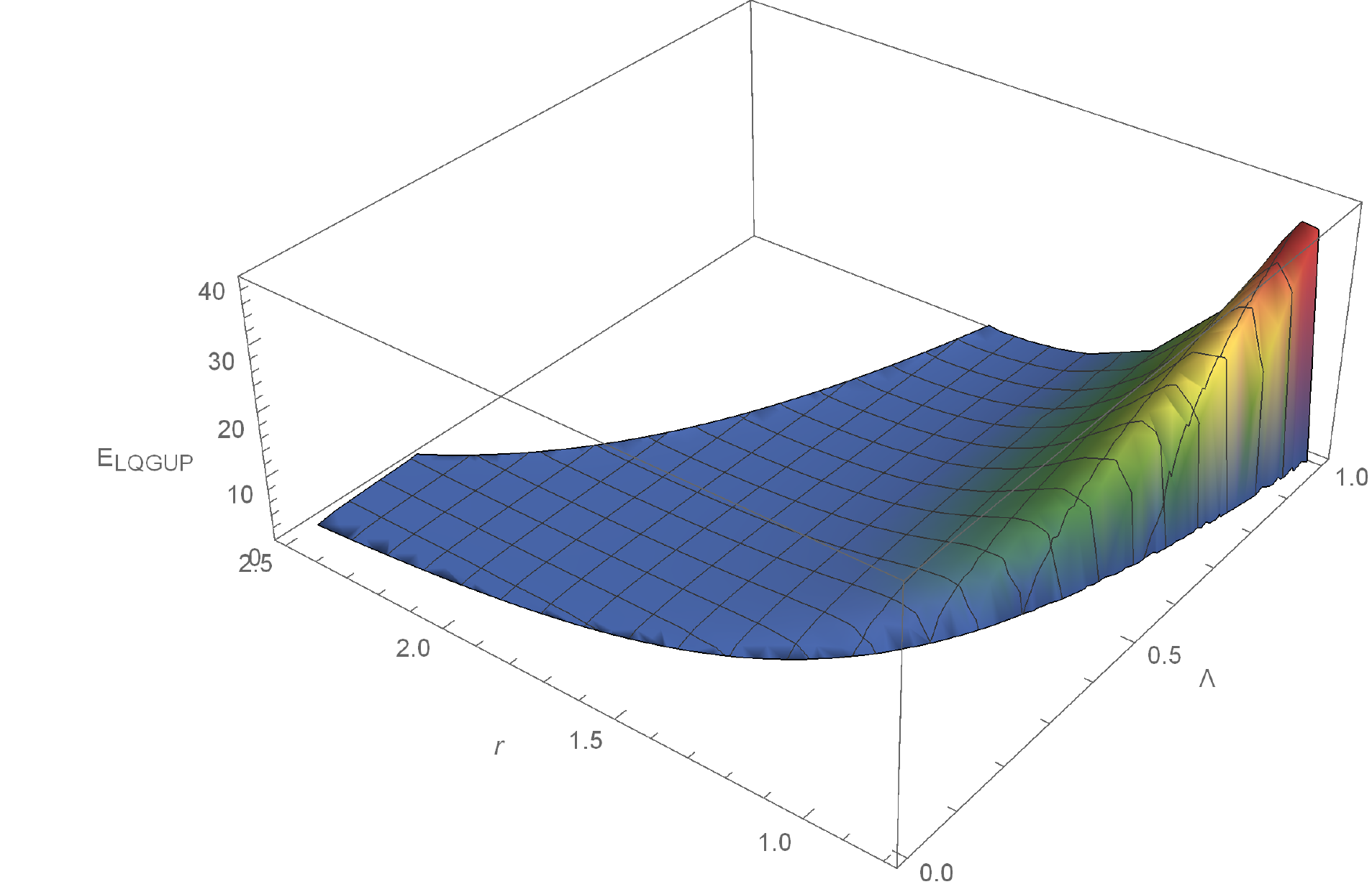}
\subcaption{ $\varpi=s=2, Q=1$.}
\end{minipage}

\begin{minipage}[t]{0.45\linewidth}
\captionsetup{justification=centering}
\includegraphics[width=\linewidth]{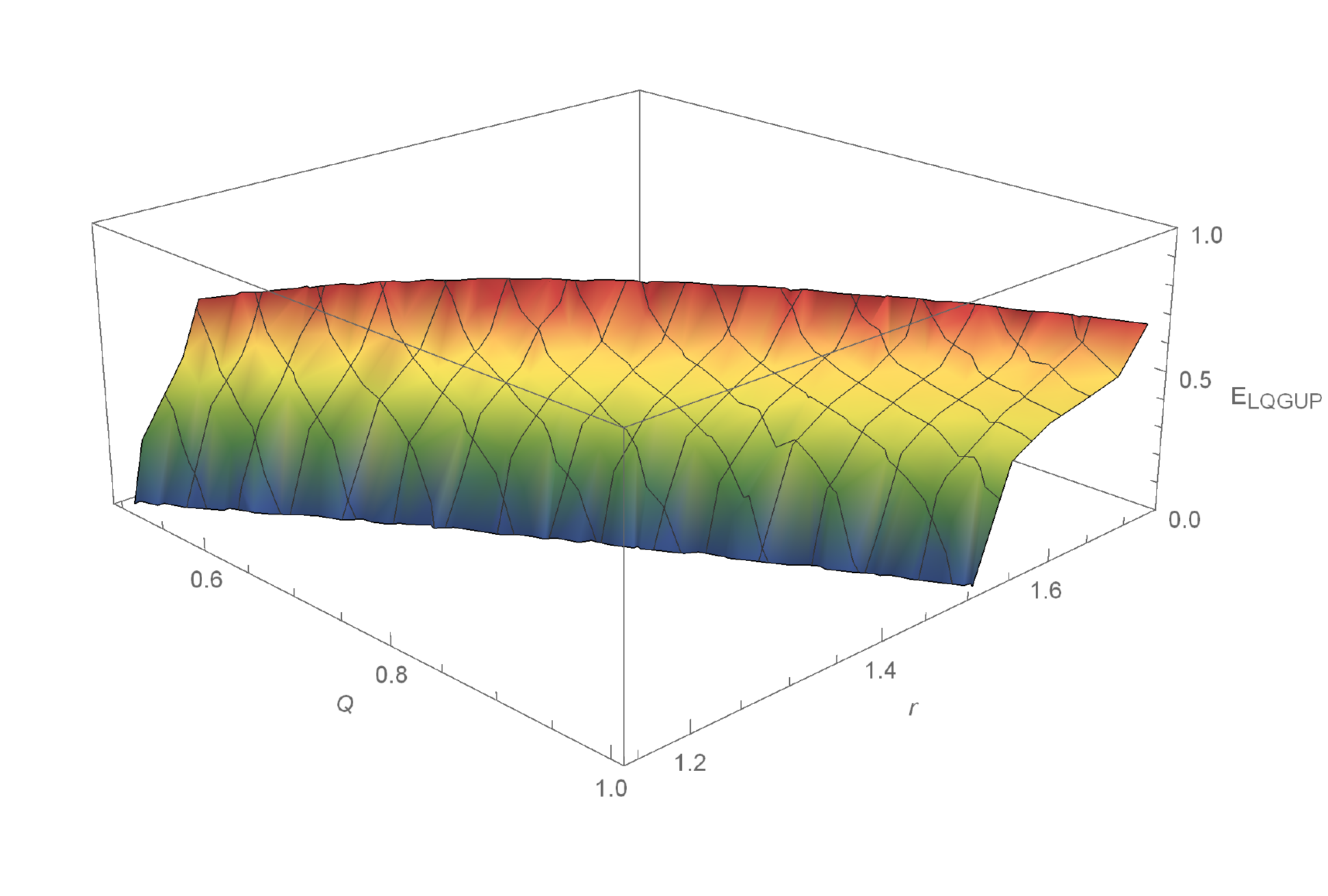}
\subcaption{ $\varpi=-(s=2), \Lambda=1$.}
\end{minipage}\hfill
\begin{minipage}[t]{0.45\linewidth}
\captionsetup{justification=centering}
\includegraphics[width=\linewidth]{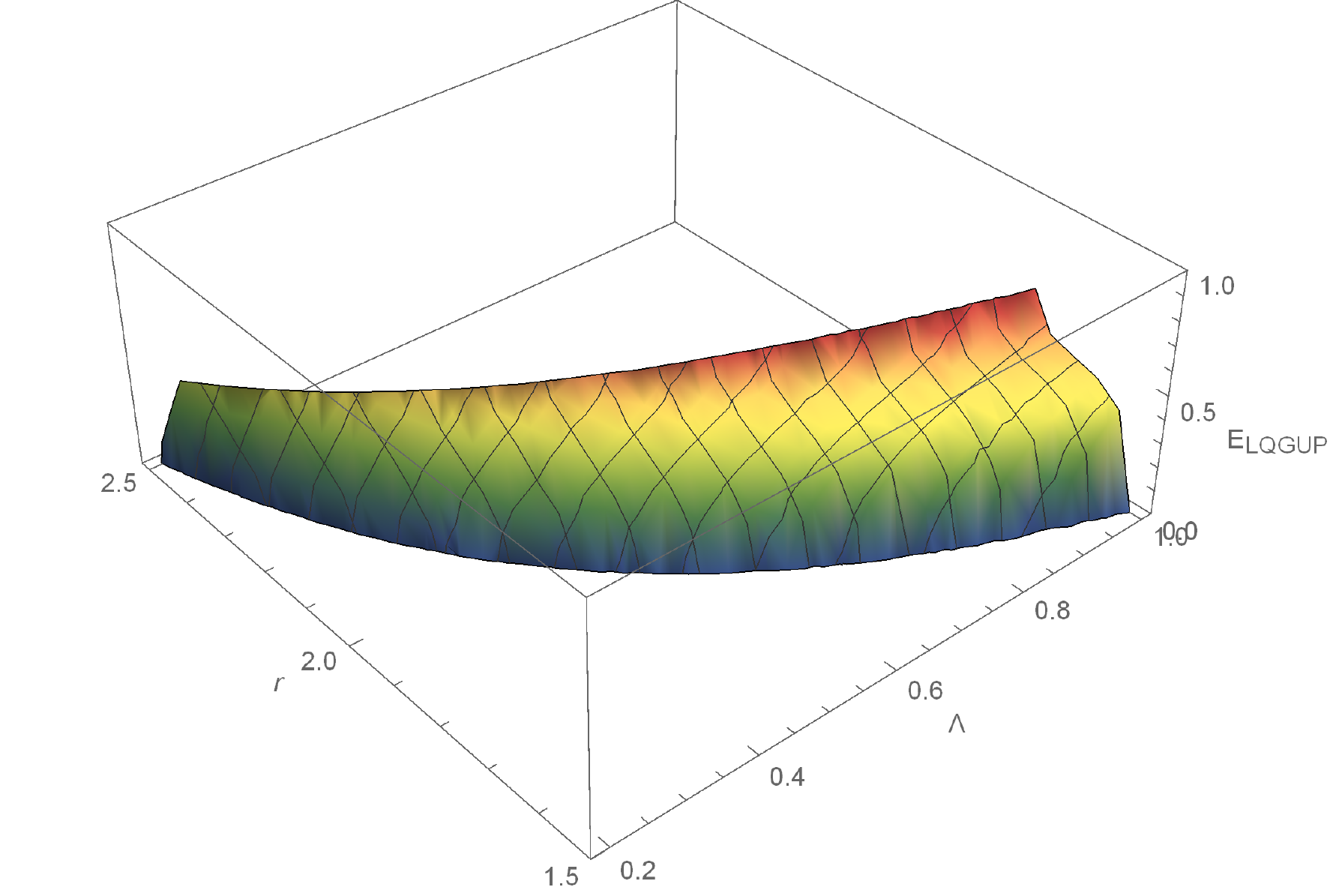}
\subcaption{ $\varpi=-(s=2), Q=1$.}
\end{minipage}

\begin{minipage}[t]{0.45\linewidth}
\captionsetup{justification=centering}
\includegraphics[width=\linewidth]{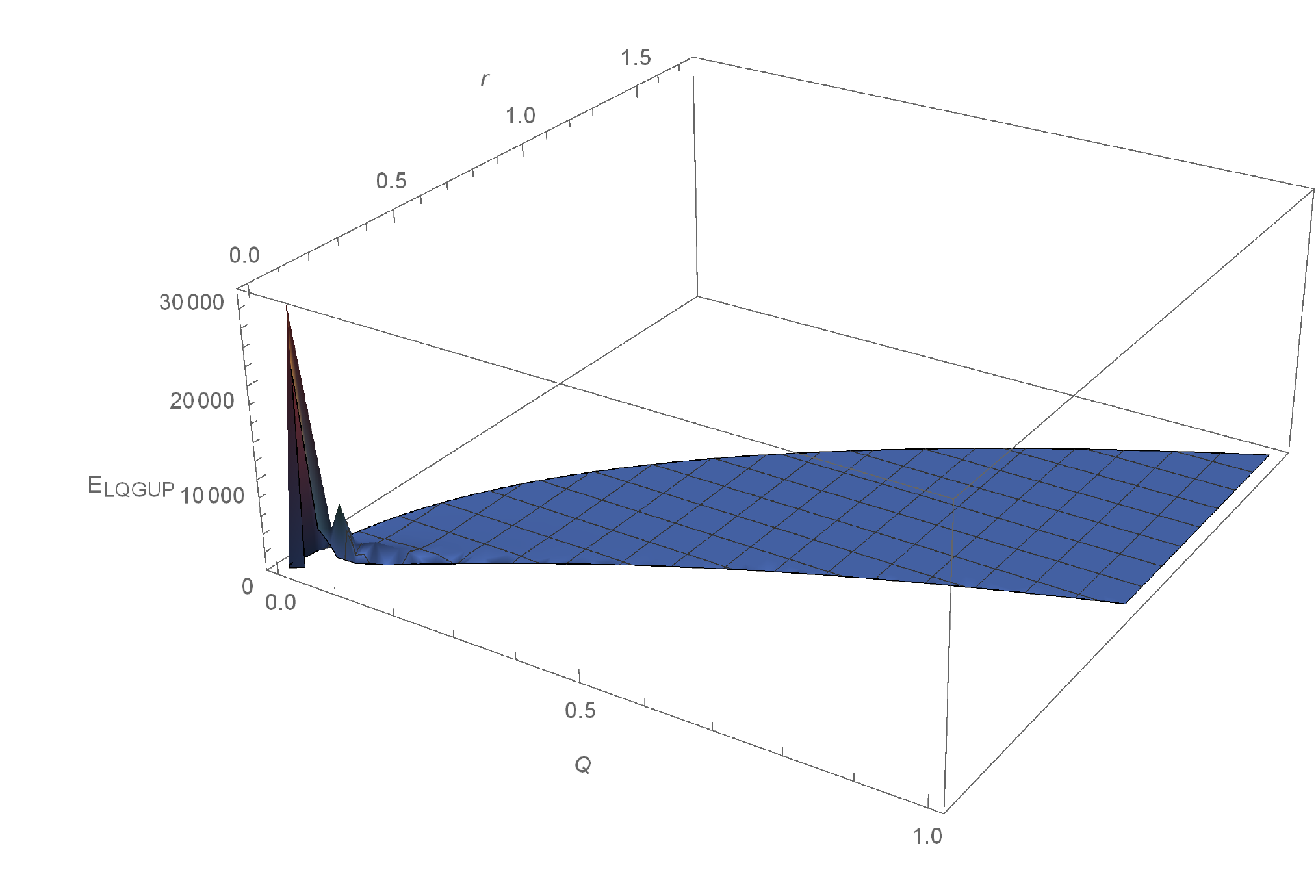}
\subcaption{ $\varpi=s=-2, \Lambda=1$.}
\end{minipage}\hfill
\begin{minipage}[t]{0.45\linewidth}
\captionsetup{justification=centering}
\includegraphics[width=\linewidth]{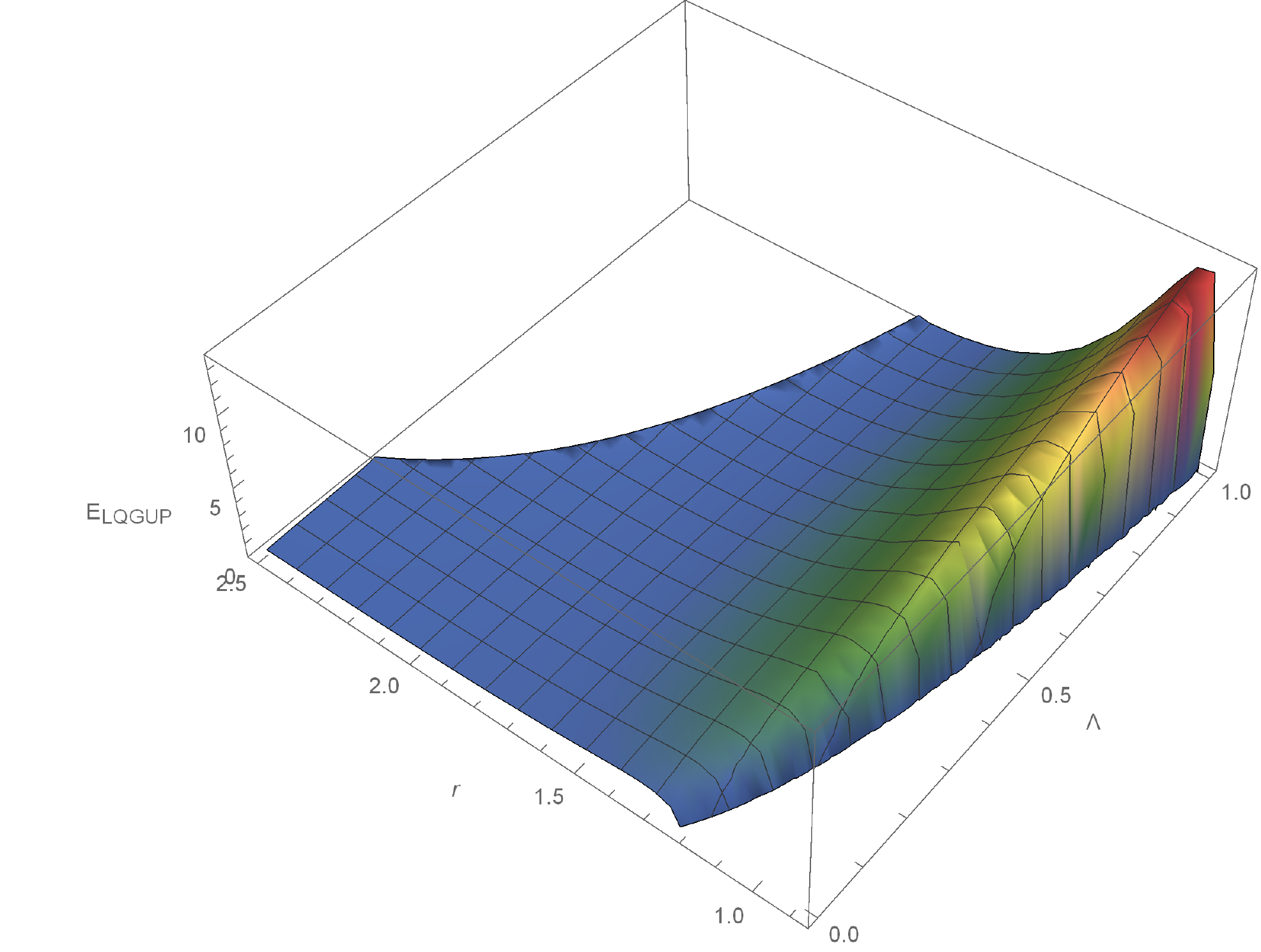}
\subcaption{ $\varpi=s=-2, Q=1$.}
\end{minipage}

\begin{minipage}[t]{0.45\linewidth}
\captionsetup{justification=centering}
\includegraphics[width=\linewidth]{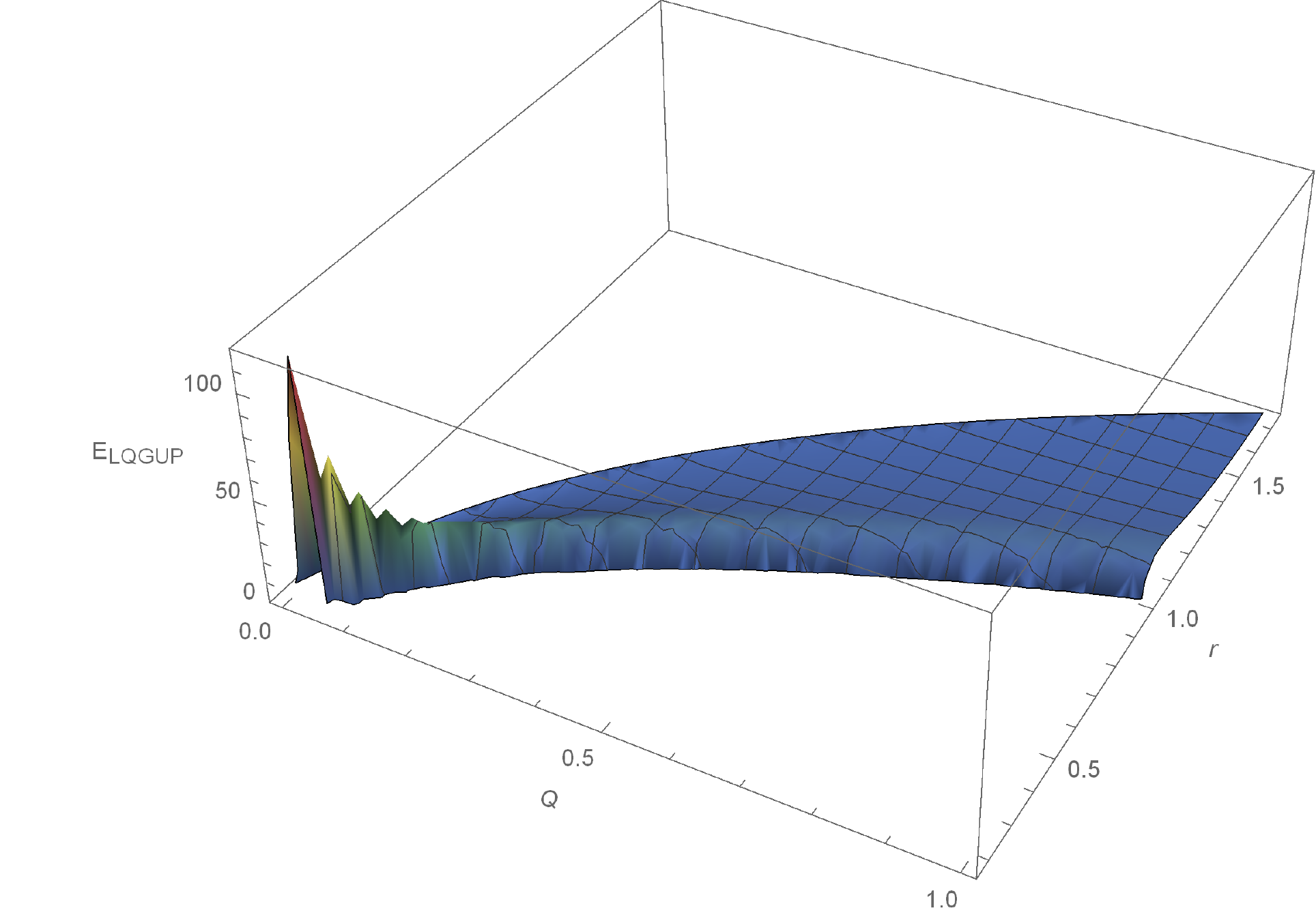}
\subcaption{ $\varpi=-(s=-2), \Lambda=1$.}
\end{minipage}\hfill
\begin{minipage}[t]{0.45\linewidth}
\captionsetup{justification=centering}
\includegraphics[width=\linewidth]{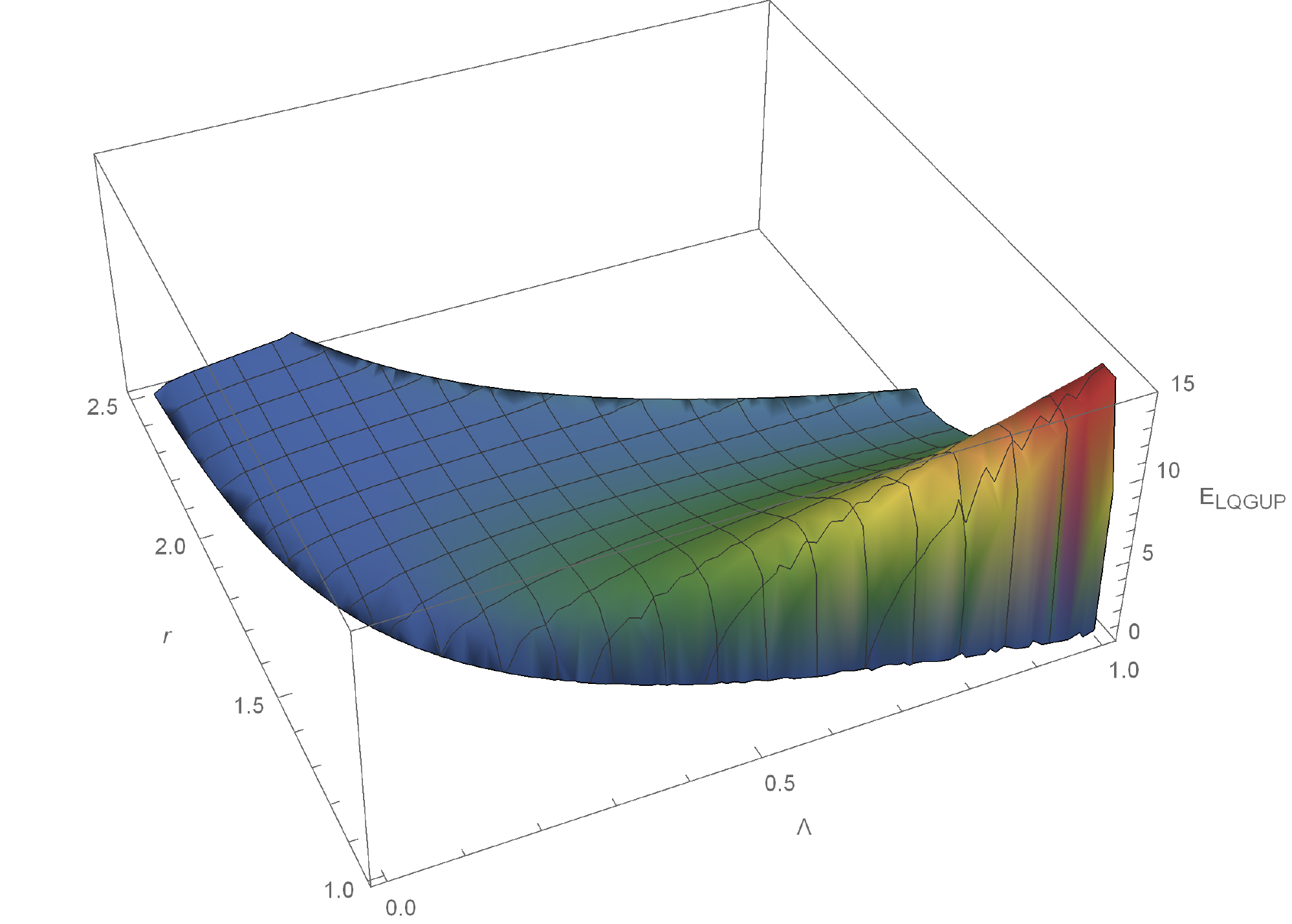}
\subcaption{ $\varpi=-(s=-2), Q=1$.}
\end{minipage}
\captionsetup{justification=centering}
\caption{Tunneling energy $E_{\text{LQGUP}}$ for $\varpi=\pm (s=\pm 2)$ bosonic massless fields.}
\label{fig.5}
\end{figure}

\newpage

\end{arabicfootnotes}
\end{document}